\documentclass[format=acmsmall, review=false, screen=true, authorversion]{acmart}

\setcitestyle{numbers,sort&compress}
\usepackage{booktabs} 

\usepackage[ruled]{algorithm2e} 

\acmJournal{TOSN}
\acmVolume{15}
\acmNumber{4}
\acmArticle{47}
\acmYear{2019}
\acmMonth{10}

\setcopyright{acmlicensed}

\acmDOI{10.1145/3356341}

\usepackage{balance}
\usepackage{comment}
\usepackage{csquotes}

\usepackage{microtype}
\usepackage{xspace}

\usepackage{stfloats}
\usepackage{multirow}
\usepackage{array}
\usepackage[export]{adjustbox}
\usepackage{graphicx}
\usepackage{subfig}
\usepackage{colortbl}

\usepackage{footnote}
\makesavenoteenv{tabular}
\makesavenoteenv{table}
\usepackage{footmisc}
\usepackage{url}
\newcommand{\name}{Whisper\xspace}
\newcommand{\Packet}{Signaling packet\xspace}
\newcommand{\packet}{signaling packet\xspace}
\newcommand{\packets}{signaling packets\xspace}
\newcommand{\packlet}{packlet\xspace}
\newcommand{\packlets}{packlets\xspace}

\newcommand{\glossy}{Glossy\xspace}

\newcommand{\celll}[3]{
	\begin{minipage}[#1]{#2}
		\flushleft
		#3
	\end{minipage}
}

\begin{document}
	\title[Fast Flooding for Low-Power Wireless Networks]{\name: Fast Flooding for Low-Power Wireless Networks}
	
	\author{Martina Brachmann}
	\affiliation{%
		\department{Networked Embedded Systems}
		\institution{RISE Research Institutes of Sweden}
		\city{Stockholm}
		\country{Sweden}
	}
	\email{martina.brachmann@ri.se}
	
	\author{Olaf Landsiedel}
	\affiliation{%
		\department{Distributed Systems}
		\institution{Kiel University}
		\city{Kiel}
		\country{Germany \&}
		\institution{Chalmers University}
		\city{Gothenburg}
		\country{Sweden}
	}
	\email{ol@informatik.uni-kiel.de}
	
	\author{Diana G\"{o}hringer}
	\affiliation{%
		\department{Adaptive Dynamic Systems}
		\institution{Technische Universit\"{a}t Dresden}
		\city{Dresden}
		\country{Germany}
	}
	\email{diana.goehringer@tu-dresden.de}
	
	\author{Silvia Santini}
	\affiliation{%
		\department{Faculty of Informatics}
		\institution{Universit\`{a} della Svizzera italiana (USI)}
		\city{Lugano}
		\country{Switzerland}
	}
	\email{silvia.santini@usi.ch}
	
	\begin{abstract}
		This paper presents \name, a fast and reliable protocol to flood small amounts of data into a multi-hop network.
		\name makes use of synchronous transmissions, a technique first introduced by the Glossy flooding protocol.
		In contrast to Glossy, \name does not let the radio switch from receive to transmit mode between messages.
		Instead, it makes nodes continuously transmit identical copies of the message and eliminates the gaps between subsequent transmissions.
		To this end, \name embeds the message to be flooded into a \textit{\packet} that is composed of multiple \emph{\packlets} -- where a \packlet is a portion of the message payload that mimics the structure of an actual packet.
		A node must intercept only one of the \packlets to detect that there is an ongoing transmission and that it should start forwarding the message. 
		This allows \name to speed up the propagation of the flood, and thus, to reduce the overall radio-on time of the nodes.
		Our evaluation on the FlockLab testbed shows that \name 
		achieves comparable reliability but 2x lower radio-on time than 
		Glossy. 
		We further show that by embedding \name in an existing data collection application, we can more than double the lifetime of the network.

	\end{abstract}

	%
	%

	
	%
	%
	

	\maketitle
	%
	
\section{Introduction}
\label{sec:intro}

Many application scenarios for low-power wireless networks require the reliable and fast exchange of small data values. Examples include the dissemination of parameters  -- e.g., the set-point of a valve -- in industrial control systems or smart buildings. 
In these settings, the communication is either periodic or event-driven. 
Approaches like Glossy~\cite{Ferrari2011+Glossy} or LWB~\cite{Ferrari2012+LWB} can be used to  reliably and quickly share data periodically in a low-power wireless network. 
Crystal~\cite{Istomin2016+Crystal,Istomin2018+Crystal2} and other approaches~\cite{Sutton2015+Zippy, Istomin2018+Crystal2, Lim2017+Competition, Escobar2018+Competition, Trobinger2018+Competition, Ma2018+Competition, Nahas2018+Competition} extend Glossy to event-driven systems by periodically sampling the network for potential~events.

In this paper, we present \name, a novel communication primitive that addresses the challenge of quickly and efficiently sharing small amounts of information in either periodic or event-based application scenarios.  
\name relies on synchronous transmissions to provide fast and energy-efficient communication in low-power wireless multi-hop networks. 
First, \name integrates the flood -- consisting of multiple replicated packets as in Glossy -- into a single packet transmission. 
This reduces the duration of a flood, eliminates gaps, i.e, the radio turnaround time, and thereby allows \name to reduce radio-on time and energy consumption.
Second, its compact flooding format and the lack of gaps between packets enable \name to introduce energy-efficient sampling strategies, i.e, the ability to sample the channel for a packet and not, as done in Glossy, LWB, Crystal and most others protocols to rely on idle listening of the radio.
    
\name can be used as \enquote{standalone} dissemination protocol for small data and as service for other protocols, e.g., as an efficient wake-up primitive to reduce the protocol's idle listening overhead. 
The latter is the case in the event-based transmission of large data packets. 
In such scenarios, the nodes listen to potential packets for a predefined interval before turning the radio off.
Using \glossy-like protocols, the length of this interval depends on the packet size, the number of hops in the network, and the number of packet repetitions~$N_{tx}$. 
For example, it would take \glossy almost 67~ms to disseminate a 127~byte packet in a 6-hop network with $N_{tx} = 3$. 
Thus, all nodes must keep the radio turned on for this amount of time, even when no packet is disseminated.
Using \name, the nodes keep their radio on for at most a \name slot, which is less than 3~ms in a 6-hop network. 
Only if they have received a packet during the Whisper slot, they 
await the actual data packet afterward. 
Otherwise, they keep their radio turned off. 
Thus, using \name, nodes only communicate when they have an event to share. 

In summary, \name is designed around three cornerstones:
\begin{itemize}	
\item \name compacts the network flood into a single packet without \enquote{gaps} to reduce latency and radio-on time and to enable efficient sampling.
\item \name employs strategies for sampling so that nodes can 
energy-efficiently determine whether there is data to exchange during a round or not.
\item \name exploits synchronous transmissions for fast and reliable flooding.
\end{itemize}

We show the efficiency of \name as primitive for fast and 
reliable network-wide flooding of small amounts of 
data by evaluating it on the publicly available FlockLab testbed.
For example, in \name nodes can determine whether a flood is ongoing 
within less than 3~milliseconds in a 6-hop network, whereas in Crystal 
and in several solutions of the EWSN dependability competitions, 
nodes are awake for the duration of the network flood, which on 
FlockLab, for example, commonly takes 5~milliseconds. 
We further show that \name has an up to 50\% higher network 
lifetime 
than Glossy and it can increase the network lifetime of Crystal by a 
factor of 2.3 when used as wake-up service within Crystal.

The remainder of this paper is structured as follows.
We present a high-level overview of \name in 
Sec.~\ref{sec:overview} 
and discuss relevant details of its design in 
Sec.~\ref{sec:details}.
Using our implementation for TelosB nodes in Contiki, we compare 
\name in various dissemination scenarios with \glossy in  
Sec.~\ref{sec:eval} and use \name as wake-up primitive in Crystal in 
Sec.~\ref{sec:eval_crystal}.
A discussion of related work follows in Sec.~\ref{sec:related_work} 
and Sec.~\ref{sec:conclusion} concludes the paper.
\section{\name: How it works}
\label{sec:overview}

\begin{figure}[htbp]
    \centering
        \includegraphics[width=\linewidth]{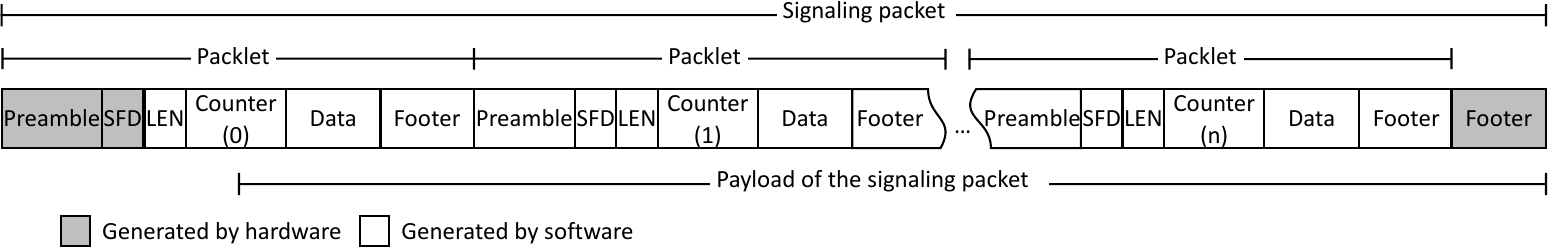}
    \caption{\emph{Format of a \packet.} The gray-shadowed footer at the end of the packet is created only when the radio is used in buffered mode.}
        \label{fig:packet_format}
\end{figure}

At its core, \name is a communication primitive that allows to quickly and reliably flood data into a multi-hop network. 
In the following, we describe the three building blocks of 
\name's design: \packlets, direction-aware channel sampling, and 
synchronous transmissions. 

\paragraph{\Packet and packlets} 

\begin{figure}
	\centering
	\subfloat[\emph{\name.}]{%
		\includegraphics[width=.9\linewidth]{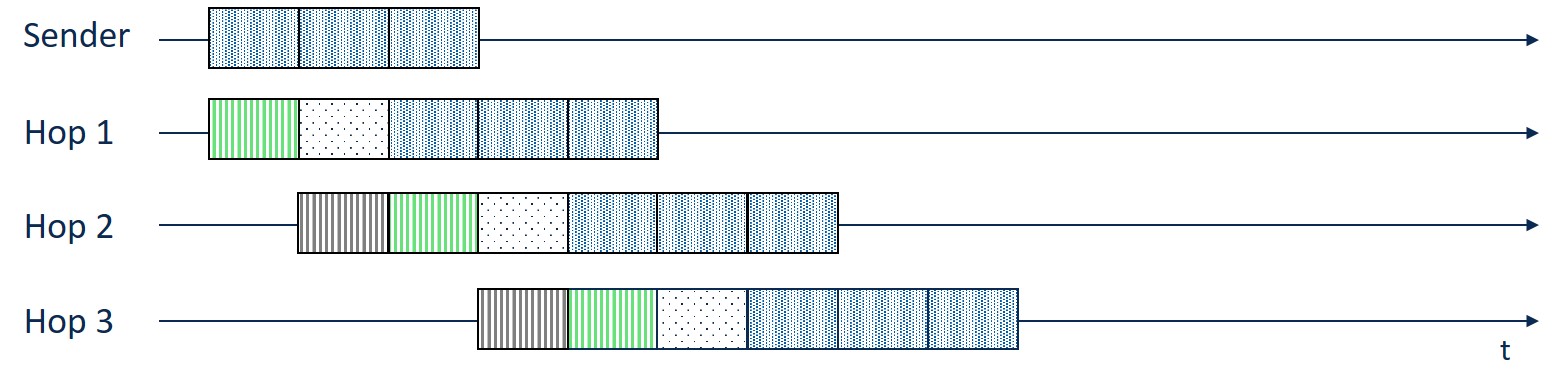}
		\label{fig:overview_whisper}
	}

	\subfloat[\emph{\name with lazy sampling.}]{%
		\includegraphics[width=.9\linewidth]{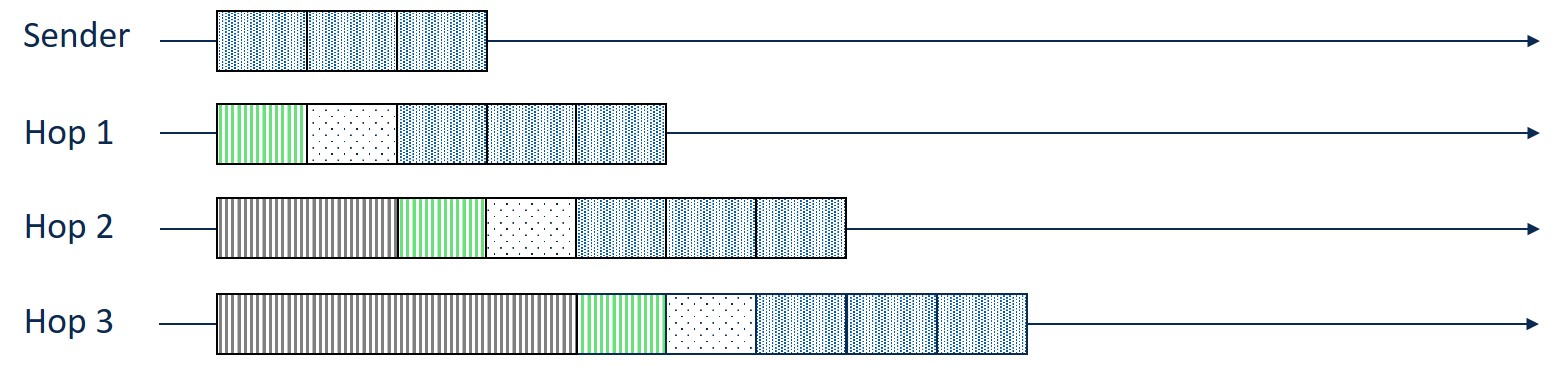}
		\label{fig:overview_whisper_lazy}
	}
    
    \subfloat[\emph{Glossy.}]{%
		\includegraphics[width=.9\linewidth]{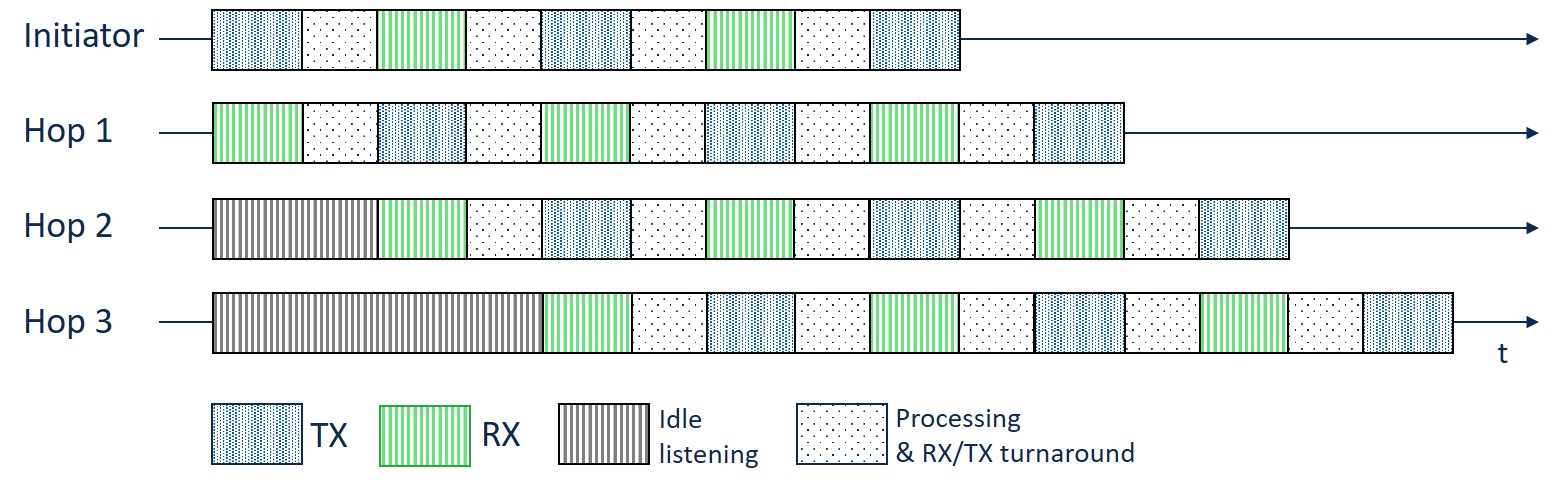}
		\label{fig:overview_glossy}
	}
	\caption{\emph{\name eliminates gaps.} The time needed to switch the radio between receive (RX) and transmit (TX) mode (and vice versa) causes communication ``gaps'' to occur in Glossy. By transmitting a log train of \packlets, \name does not need to perform the RX/TX turnaround and thus eliminates these gaps.}
	\label{fig:overview}
\end{figure}

In \name, a node that needs to send data -- hereafter referred to as the \textit{sender} -- transmits a \emph{\packet}, which looks as depicted in Fig.~\ref{fig:packet_format}.
It consists of several\footnote{While the number $N_{tx}$ of 
\packlets included in a \packet is a configurable parameter, our 
results show that a default value of $N_{tx}=3$ is sufficient to 
achieve very high reliability.} \emph{\packlets}, whereas a \packlet 
is a piece of message payload that has the structure of an actual 
IEEE~802.15.4 packet, including preamble, start-of-frame 
delimiter~(SFD) and footer.
\name's \packet, thus, mimics a train of short, identical packets being sent continuously by the radio, as illustrated in Fig.~\ref{fig:overview_whisper}. 

This is a core difference between \name and, e.g., Glossy \cite{Ferrari2011+Glossy}. 
In \glossy\xspace -- and in all protocols that build upon it -- the 
nodes 
continuously switch between sending and receiving mode, and thus, 
leave ``gaps'' between two transmissions, as depicted in 
Fig.~\ref{fig:overview_glossy}. 
The absence of such gaps significantly increases the speed at which 
the network flood can propagate and, thus, it reduces the time nodes 
must keep their radio on.

It is also worth noting that there exist other techniques, 
e.g.,~\cite{Lim2017+Competition,Escobar2018+Competition,Ma2018+Competition},
 that modified Glossy to make it transmit packets consecutively 
instead of alternating between reception and transmission.
These existing approaches, however, still suffer from the existence of the gaps because the radio transceiver must perform the RX/TX turnaround each time after packet transmission completes. 
By embedding \packlets in the payload of a single \packet, \name 
completely eliminates gaps between consecutive transmissions, because 
there is no RX/TX turnaround.

While \name can support payloads of arbitrary length, it is best 
suited to flood small amounts of data, as we also discuss in 
Sec.~\ref{subsec:payload_size}.  
Small payloads occur frequently in real scenarios, e.g., when a configuration parameter, which may be coded using just a few bits, must be communicated to all nodes in a network. 
More importantly, \name can be used as a very efficient signaling 
primitive to, e.g., notify to all nodes in a network that they must 
stay awake to help forwarding incoming data packets. In this 
scenario, no data or only very little data is actually transmitted 
and thus a small payload size does not represent a limiting factor.

\begin{figure}
	\centering
		\includegraphics[width=.9\linewidth]{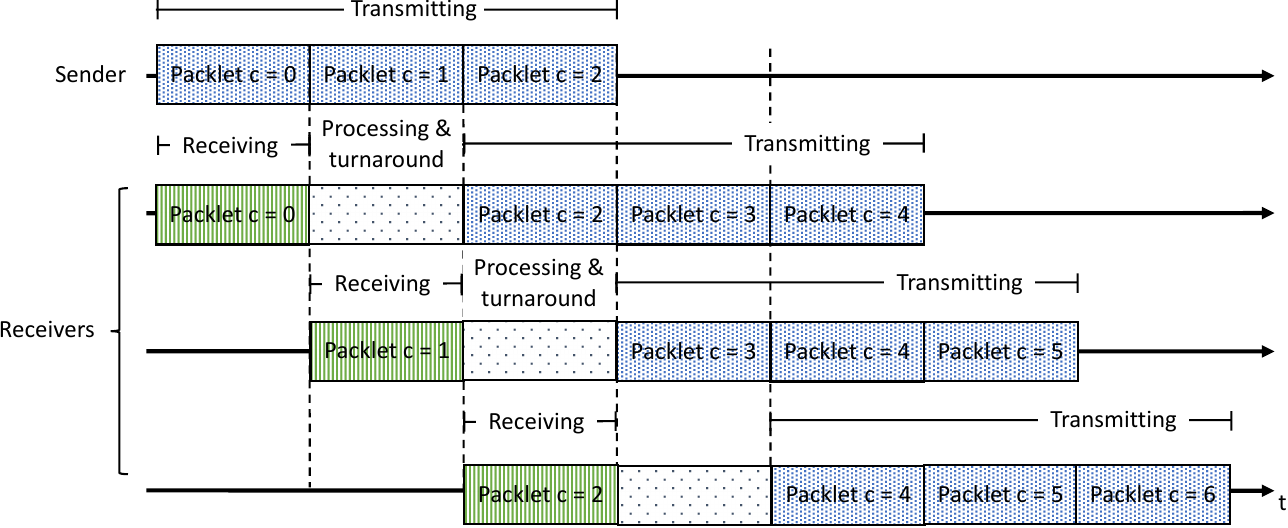}
		\label{fig:packet_transmission}
	\caption{\emph{\name's operation.} Nodes receive a \packlet, process it while turning their radio to transmit mode and afterwards transmit their \packet.}
	\label{fig:fastawake_operation}
\end{figure}

\paragraph{Sampling strategy} 
For nodes to be able to detect the presence of a \packet, they must 
regularly switch their radios on and check the channel for incoming 
transmissions. 
The more often this channel check is performed -- and the longer each check lasts --  the higher is the duty cycle of the nodes and thus their energy consumption. 
The possibility to design thrifty sampling strategies -- which is opened up by the use of packlets -- is thus instrumental to reduce the overall radio-on time and thus the duty cycle of nodes running \name. 

A straightforward sampling strategy -- to which we refer to as 
\textit{lazy sampling} (see Fig.~\ref{fig:overview_whisper_lazy}) -- 
consists in making all nodes switch their radios on at the beginning 
of a communication slot.
This strategy is used in Glossy and other approaches such as LWB~\cite{Ferrari2012+LWB} or Crystal~\cite{Istomin2016+Crystal} and can be used in \name too.
When adopting lazy sampling, nodes must wait for an incoming transmission long enough so that a message from the initiator can propagate through the entire network. 
This can, however, take several milliseconds in a network of few hops and represents a high cost in terms of energy consumption, especially if no packet is transmitted.  

To cope with this problem, \name uses an alternative sampling strategy -- which we dub \textit{direction-aware sampling}.
It exploits the fact that in many practical scenarios the network topology is usually fixed or changes slowly and that data traffic flows in one direction only -- e.g., from an initiator to all other nodes in a network in a data dissemination scenario or from a random node in the network to a central sink node in case of event-driven aperiodic communication.
Thus, nodes can estimate their distance in hops from the sender or 
destination, respectively, and switch on their radios only when a 
\packet is likely to ``pass-by'', as shown in 
Fig.~\ref{fig:overview_whisper}.

\paragraph{Synchronous transmissions} 
To ensure a fast and reliable propagation of the \packet, \name exploits synchronous transmissions. 
When a neighbor of the sender turns its radio on, it needs to intercept only one of the \packlets to detect the existence of a \packet. 
If no \packlet is detected, the node switches its radio off to save 
energy.
If a \packlet is instead successfully received, the node keeps its radio on and helps propagating the \packet. 
It does so by joining the ongoing synchronous transmission with its own \packet, which is again a single packet made of multiple \packlets.
To this end, a node that starts sending a \packet must ensure that its own \packlets overlap with the \packlets that are already being transmitted by other nodes.

\name takes special account of scenarios in which nodes share the 
same data, e.g., when a controller disseminates a new configuration 
parameter or when \name is used as wake-up primitive.
In these cases, \packlets sent by nearby nodes must fulfill two 
conditions: they must be identical and must be sent at almost exactly 
the same time instant\footnote{As known from 
Glossy~\cite{Ferrari2011+Glossy}, the temporal displacement between 
concurrently transmitted packets must (in IEEE 802.1.5.4) not exceed 
0.5~$\mu$s to allow for constructive interference to 
occur.}, as 
schematically illustrated in Fig.~\ref{fig:overview_whisper}. 
We discuss in Sec.~\ref{sec:details} how \name manages to fulfill 
both these conditions.

\name also supports the concurrent transmission of different data.
The procedures for sending identical data packets and different data packets are identical.
The latter case, however, relies on the capture effect instead of 
constructive interference.
\section{\name: A Closer Look}
\label{sec:details}

After having presented the core design elements of \name in the 
previous section, we now discuss in detail its most relevant features.
\subsection{The \packet}
\label{subsec:packet}
The design of the \packet as shown in Fig.~\ref{fig:packet_format} 
makes \name able to send a train of packets back-to-back, i.e., 
without any gaps between consecutive transmissions. 
We argue that this is an essential stepping stone to: (a) lower the duration of a network flood; (b) enable sampling strategies that reduce nodes' duty cycle; and (c) simplify the timing of synchronous transmissions.


\name uses the payload of the \packet to simulate several packets being sent back-to-back.
This is achieved by the use of \packlets, a technique inspired by the 
multi-header approach presented in~\cite{Liang2010+Multi-Headers}. 
As illustrated in Fig.~\ref{fig:packet_format}, a \packlet in \name 
consists of at least five fields\footnote{The \emph{data} field can 
	have a length of zero, e.g., when running \name as wake-up 
	primitive.}: a preamble, a 1-byte SFD, 1-byte length field, a 1-byte 
payload, and a 2-byte footer, which includes a Frame Check 
	Sequence (FCS).
Nonetheless, the \packlet format in \name is user-configurable and can be changed or extended to contain, e.g., larger payloads.
When a receiver starts listening for incoming packets it only needs 
to intercept a single preamble and SFD of one of the \packlets to 
detect an ongoing transmission.


While the IEEE 802.15.4-compliant length of the preamble is 4 bytes, in some radios -- e.g., the CC2420~\cite{CC24202013} -- both the preamble length and the SFD are configurable parameters. 
To reduce the total length of a \packet, and thus, decrease the 
radio-on time of the nodes, \name's default implementation sets the 
length of the preamble to 2 bytes. 
While this assumes that \name can exploit low-level features of the 
transceiver and makes it non-IEEE~802.15.4-compliant, we believe that 
it is important to explore the potential of \name's design beyond 
current technological limits. 
Several other authors have indeed explored before us 
non-IEEE~802.1.5.4-compliant techniques to design energy-efficient 
protocols~\cite{Polastre2004+BMAC, El-Hoiydi2004+WiseMac, 
	Buettner2006+XMAC}. 
Nonetheless, \name can operate with a preamble of arbitrary length 
and can thus, if required, also be used with a standard-compliant 
preamble of 4~bytes.
While a longer preamble affects performance, we show in 
Sec.~\ref{sec:eval} that \name outperforms Glossy also with a 
preamble \mbox{length of 4 bytes}.

Given the above description, a \packlet is by default 7~bytes long. 
Since IEEE~802.15.4 radios transmit at a rate of 
250~kbit/s, the transmission of one \packlet\xspace 
-- henceforth denoted with $T_{packlet}$ -- lasts 
224~$\mu$s. 
The sender sends $N_{tx}$ \packlets and its total transmission time 
with $N_{tx}=3$ is thus 672~$\mu$s. 
The radio of other nodes in the network is instead active for at 
least the duration of $N_{tx}+2$ \packlets. 
This is because, as depicted in Fig.~\ref{fig:overview_whisper}, 
before being able to send its $N_{tx}$ \packlets a node must first 
receive one \packlet and then switch the radio from receive to 
transmit mode, thereby adding the duration of 2 \packlets to the 
total time needed to transmit $N_{tx}$ \packlets.
Further, nodes must sample the channel, which further adds to the 
time they must keep the radio on.
With direction-aware sampling, for instance, nodes persist in idle 
listening for the duration of roughly one \packlet, thus, bringing 
the otal radio-on time to $(N_{tx}+2+1) \cdot 
T_{packlet}$, i.e., 1,344~ms by setting $T_{packlet} =$ 224~$\mu$s. 

In contrast, nodes running \glossy in the same scenario must keep 
their radio on for at least $2N_{tx}$ transmissions. %
This is because, as shown in Fig.~\ref{fig:overview_glossy}, a node in \glossy both receives and transmits a packet $N_{tx}$ times. 
Assuming that also \glossy sends packets with a 1-byte payload -- 
and that a \glossy packet is as long as a \packlet\xspace -- 
each node actively transmits or receives for $2N_{tx} T_{packlet}$, 
i.e., 1.344~ms when $N_{tx} = 3$. 
\glossy must, however, also continuously switch between receive and 
transmit mode, as illustrated in Fig.~\ref{fig:overview_glossy}.
This RX/TX turnaround of the radio takes 
192~$\mu$s~\cite{IEEE802.15.4} and nodes must turn the radio from 
receive to transmit mode $N_{tx}-1$ times, which adds almost 1~ms 
(i.e., 960~$\mu$s) of additional radio on time. 
The radio-on time of \glossy is thus almost twice as long as that of 
\name (2,304~ms vs. 1,344~ms) -- even though we did not account for 
the time spent in idle listening by nodes running \glossy nor for 
\glossy's software delay, which should be added to the RX/TX 
turnaround time.
We also did not consider -- neither in the calculation above nor in Fig.~\ref{fig:overview} -- the guard times that are present in both \name and \glossy. 
A guard interval is usually short\footnote{The reference 
	implementation of \glossy we use in the evaluation has a guard time 
	of roughly 130~$\mu$s (measured experimentally). 
	In~\cite{Istomin2016+Crystal}, Istomin et al. showed that a guard 
	time of 150$~\mu$s is sufficient to compensate for clock drifts that 
	accumulate over 5~minutes.} and appears only once at the beginning of 
the idle listening phase. 
It, thus, has only little influence on the computation presented 
above.


This back-of-the-envelope calculation shows that the superior performance of \name with respect to \glossy\xspace -- discussed in detail in Sec.~\ref{sec:eval_compare_glossy} -- is mainly due to the fact that \name eliminates the gaps between consecutive transmissions.
This advantage persists even if \name is used with lazy sampling, as in this case the time spent in idle listening is roughly the same as for Glossy. 

In the example discussed above, we assume \glossy packets with a 
payload of 1 byte. 
The payload of standard \glossy packets is, however, 4 bytes: a 2 
byte sequence number, a 1 byte \glossy header, and a 1 byte relay 
counter~\cite{Ferrari2011+Glossy}. 
We reduce the payload size to 1 byte (we keep only the field \textit{relay\_counter}), to avoid penalizing \glossy due to its larger payload size. 
This also allows us to show -- in Sec.~\ref{sec:eval} -- that shortening the payload size in \glossy is not sufficient to make it more efficient than \name. 

\subsection{Sending identical packlets}
\label{subsec:packlets_identical}
\name natively takes special account for scenarios in which nodes 
share identical data.
For these cases, sending identical packets is a necessary 
condition 
for constructive interference to occur (or more precisely for packets 
not to interfere destructively) when synchronous transmissions are 
used. 
In \name, this translates in ensuring that all \packlets sent at the same time are identical.

The only value that changes across different packlets in the same 
\packet is the counter~$c$, which is the 1-byte payload of each 
\packlet.
As illustrated in Fig.~\ref{fig:packet_format}, the $1$st \packlet 
has counter $c=0$, the second $c=1$, and so on.
When nodes start sending their own \packet, they must 
properly set the 
value of the counter $c$ of the \packlets.
In particular -- as also shown in Fig.~\ref{fig:fastawake_operation} 
-- \name makes a node that receives a \packlet with counter $c=i$ set 
the counter of its first packlet to $c=i+2$. 
This is because while the $c=i+1$th \packlet is being transmitted, 
the node performs the RX/TX turnaround of the radio. 
In this time frame the node ``misses'' a \packlet and must wait 
until the next one starts being sent before sending its own \packet. 

Besides ensuring the values of the counter $c$ is identical for all 
synchronously transmitted \packlets, \name must also 
properly set the 
\textit{length} field of the \packlets.
This field specifies the length in bytes of the payload and the footer. 
For \packlets that have a 1-byte payload, the length field must, 
thus, be set to three. 
This can be easily done for all \packlets but the first. 
As highlighted in gray in 
Fig.~\ref{fig:packet_format}, the \packlet with counter 
$c=0$ ``borrows'' the preamble, SFD, and length field of the \packet. 
The first byte of the \packet, however, specifies how many 
bytes the radio must send before automatically ceasing to transmit.
If \name would use this mode of operation (called \textit{buffered mode} in the CC2420~\cite{CC24202013}), the (first) length field in the \packet would indicate the total length of the \packet in bytes -- which is different than three. 
This would cause the length field of each \packet to collide with the 
length field of synchronously sent \packlets.  
As a consequence, destructive interference would occur and 
result in packet drops. 

To avoid this problem, we exploit an alternative transmit mode available on certain IEEE 802.15.4 radio transceivers (including the CC2420~\cite{CC24202013}): the \textit{TXFIFO looping mode}.  
When set in this mode the radio ignores the length field and just continuously reads data from the radio buffer and transmits it. 
Once the content of the buffer has been sent, the radio wraps around and starts to read and send the data from the beginning.
This continues indefinitely until a timeout explicitly stops the transmission. 
Since the value of the length field is ignored when the radio operates in TXFIFO looping mode, \name can set the first length field to the length of a packlet -- instead that to the length of the \packet --  thus completely overcoming the problem described above. 

While this mode of operation may not be available on all IEEE 802.15.4 transceivers, which limits the portability of \name, we believe that it is important to explore novel design ideas notwithstanding current technological limits and protocol standards.
We further plan to explore an alternative approach to avoid the 
TXFIFO looping mode: using byte-wise transmission power control to 
send the length field using the smallest possible transmit 
power~\cite{Saha2017+Many-to-one}. 
When evaluating the performance of \name in 
Sec.~\ref{sec:eval}, we, nonetheless, explicitly consider a fully IEEE 802.15.4-compliant 
version of the protocol, called \name (compliant), which we describe 
in Sec.~\ref{subsec:whisper_compliant}.  

Further, \name is not limited for sharing identical data. 
Related work has already shown the efficiency of synchronously transmitting different data~\cite{Landsiedel2013+Chaos,Istomin2016+Crystal,Brachmann2016+LaneFlood,Yuan2014+Sparkle} by relying on the capture effect.
Letting the nodes flood different data with \name, indeed, also relaxes the need for support of the TXFIFO looping mode.
The advantage of \name compared to existing solutions is, still, the absence of gaps resulting in a short flooding duration and thus, efficient sampling.

\subsection{Sending packlets synchronously}
\label{subsec:packlets_synchronously}

In the previous subsection we mentioned that after receiving a packlet (with counter value $c_i$), a node must wait for an entire $T_{packlet}$ before sending its first packlet (with counter $c_i+2$). 
This is because while packlet $c_i+1$ is on the air, the node must 
perform the RX/TX turnaround of the radio, which lasts $T_{turn} = 
192$~$\mu$s for IEEE~802.15.4 radios~\cite{IEEE802.15.4}.
This leaves a wait time $T_{wait} = T_{packlet} - T_{turn} - T_d$, whereas $T_d$ is the time that elapses between the rising SFD edge of a sender during transmission and the corresponding rising SFD edge of a receiver during reception.

The existence of $T_d$ is due to the fact that the reception of a pack(l)et lasts slightly longer than its transmission.
This \textit{data delay} is a common phenomenon in wireless radios, 
and each transceiver has a specific latency of the RX and TX paths, 
which is reported in the data sheets.
If not compensated for, the existence of $T_d$ would make nodes start sending the next \packlet before receivers have completed the reception of the previous one.
Including propagation delay, $T_d$ is reported to be $3 < T_d 
\leqslant 3.6$~$\mu$s~\cite{CC24202013, Yuan2013+Talk, 
	Konig2016+Maintaining-CI} and is, thus, non-negligible.
In \name, we set $T_d=3$~$\mu$s.

The existence of this fixed wait time is a further difference between \name and \glossy.
Indeed, \glossy aims at re-sending a packet as quickly as possible after receiving it (i.e., immediately after the RX/TX turnaround).
This is because the longer nodes wait to retransmit a packet, the 
stronger MCU clock instabilities become relevant and can, thus, cause 
transmissions of different nodes to 
misalign~\cite{Ferrari2011+Glossy}. 
This so-called \emph{software delay} in \glossy is $23$~$\mu$s.  
In \name, if we assume a payload of 1~byte and a preamble 
of 2 bytes,
then $T_{packlet}=224$~$\mu$s and, thus, 
$T_{wait}=29$~$\mu$s.
Although the values of $T_{wait}$ and of \glossy's software delay are relatively close to each other, the latter does not depend on the length of the packet whereas it does in the case of \name.
The wait time is, thus, more critical for \name than for \glossy. 

To cope with this issue we employ Flock, a recently presented clock compensation approach for low-power nodes~\cite{Brachmann2017+Flock}. 
Flock compensates for instabilities in the digitally controlled oscillator (DCO) that drives the MCU of many low-power hardware platforms. 
It, thus, makes \name able to ensure that packlet transmissions align within the 0.5~$\mu$s window notwithstanding the existence of $T_{wait}$ and even if $T_{wait}$ is significantly longer than 29~$\mu$s.
This, in turn, makes \name robust even with long \packlets and in challenging environments with highly unstable DCO clocks, like those considered in~\cite{Boano2014+TempLab}.

\subsection{Lazy sampling}
\label{subsec:lazy_sampling}

A straightforward way to maximize the probability that a node 
intercepts a \packet, even in scenarios with high node mobility, 
consists in making the node keep its radio in idle 
listening for the 
entire duration of a \textit{\name slot}. 
This is the time interval during which a \name flood is executed and during which -- like in \glossy and other protocols based on synchronous transmissions -- all other application tasks executing the hosting platform are suspended.

The length of the slot -- indicated as $T_{slot}$ -- is a protocol parameter and should be set depending on the expected network diameter. 
In particular, it holds: 
\begin{equation}
T_{slot} = (2d_{net} + N_{tx}) \cdot T_{packlet},
\label{eq:Tslot}
\end{equation}
where $d_{net}$ is the network diameter and $T_{packlet}$ is the time 
needed to send a packlet. 
The first addend in Eq.~\ref{eq:Tslot} accounts for the fact that \name progress at a ``speed'' of 2 packlets per hop, as illustrated in Fig.~\ref{fig:overview_whisper}.
The second addend instead considers that at the last hop, after the first packlet has been transmitted a node must still transmit $N_{tx}-1$ packlets. 
If \name is used in a network of 6 hops and with $N_{tx}=3$ and $T_{packlet}$ = 224 $\mu$s (1 byte payload), a slot length of 4,48 ms would be sufficient.
In practical settings, however, it is recommendable to use a slightly larger value to account for synchronization drifts and other issues.
In the experiments presented in Sec.~\ref{sec:eval}, for instance, we 
use $T_{slot} = 5$~ms. 

Irrespectively of the type of sampling used, once a \name slot ends nodes schedule their next wake-up according to the needs of the application. 
If, for instance, it must be checked every $5$~minutes if there is an 
update by the initiator, nodes will reschedule their wake-up 
accordingly at a time instant $t^*_{start}$ that is 5 minutes away 
from the beginning of the slot. 
To account for possible synchronization errors, \name uses as in \glossy a \textit{guard time} -- indicated as $T_{guard}$ -- and makes the node actually switch their radio on at $t^*_{start} - T_{guard}$. 
\subsection{Direction-aware sampling}
\label{subsec:direction_aware_sampling}

A significant drawback of the lazy sampling strategy sketched above is that it causes all nodes in the network to stay in idle listening for an entire \name slot -- even when no \packet is sent.
To reduce this idle listening time and thus the overall radio-on time, \name exploits a different strategy, which we call \textit{direction-aware sampling}.

The main idea behind this strategy is to let the nodes switch their radio on only shortly before the flood is expected to ``pass by''.
In low-power networks, traffic often flows in one direction only, e.g., from an initiator towards all other nodes in the network in data dissemination scenarios~\cite{Ferrari2011+Glossy, Doddavenkatappa2013+Splash, Du2015+Pando} or from all nodes to a sink in data collection~\cite{Istomin2016+Crystal}. 
If the direction of the traffic is known -- hence the name \textit{direction-aware} sampling -- \name can exploit this information to run an efficient sampling~strategy. 

In the scenario considered in \glossy, for instance, traffic always 
flows from a fixed initiator to all other nodes. 
If \name is used in this scenario, the counter~$c$ of the \packlet 
received by a forwarding node depends on the distance in hops between 
the node and the initiator. 
If the topology of the network can be assumed to be static or vary slowly, this distance -- and thus, the counter~$c$ -- can also be assumed to be constant or to vary only a little across consecutive floods.
\name exploits this situation and lets each node keep in memory two values -- $c_{min}$ and $c_{max}$ -- which are estimates of the counters 
of the \packlets with the lowest and highest counter, and thus, the 
``earliest'' and ``latest'' \packlet a node is expected to receive.
Both values $c_{min}$ and $c_{max}$ are used by the nodes to compute 
when to turn the radio on and off, respectively.
The value of $c_{min}$ is set to the lowest value of $c$ ever received. 
The value for $c_{max}$ can be computed as follows:

\begin{equation}
c_{max} = (c_{max} + c)/2   \quad \text{if: } c \ge c_{max}-2
\label{eq:c_max}
\end{equation}

The term $c_{max}-2$ in the condition of Eq.~\ref{eq:c_max} accounts 
for the progression speed of $2$ in Whisper. 
Thus, Eq.~\ref{eq:c_max} allows the filtering of outliers that could 
result in a high sampling duration, and thus, high energy 
consumption, while also being able to react to changing channel 
conditions.
Underestimating $c_{max}$ would, thus, cause a node to switch off its 
radio too early, which in the worst case could stop the propagation 
of the flood. 
The strategy chosen to set both $c_{min}$ and $c_{max}$ are very 
conservative and can definitely be improved in future work. 
The design of \name actually opens up opportunities for designing further smart sampling strategies beyond the two -- lazy and direction-aware -- discussed in this paper. 



Once $c_{min}$ and $c_{max}$ are known, a node can compute the start and the duration of its sampling interval as follows:
\begin{equation}
t_{start} = t^*_{start} - T_{guard} + max(0,c_{min}-1)\cdot T_{packlet}
\label{eq:Tstart}
\end{equation}

\begin{equation}
T_{sampling} = (\lfloor c_{max} \rfloor + (N_{tx} + 1)) \cdot 
T_{packlet} - t_{start}
\end{equation}

In Eq.~\ref{eq:Tstart}, $t^*_{start}$ indicates the time at which the sender is expected to start its transmission, whereas $T_{guard}$ is the guard time that protects against possible synchronization drifts. 
When a node did not yet receive its very first \packlet, it sets $c_{min}=0$ and $T_{sampling} = T_{slot}$. When the first packlet with counter $c$ is received, the node sets $c_{min}=c_{max}=c$. Afterwards, the mechanisms mentioned above are used to update $c_{min}$ and $c_{max}$.

The discussion above assumes that \name is used in a data dissemination scenario. 
%
In Sec.~\ref{sec:eval_crystal}, we show how direction-aware sampling can be applied also in a data collection scenario.

\subsection{\name (compliant)}
\label{subsec:whisper_compliant}

The standard version of \name described above exploits low-level mechanisms of the radio transceiver. 
For the sake of completeness, we consider in our evaluation also an IEEE-802.15.4-compliant version of \name, called \name (compliant).
This version uses a 4-byte preamble and the radio in buffered mode, which has the following two consequences.
First, the first length field of the \packet must be set to the 
actual length of the payload and will, thus, collide with the length 
field of synchronously sent \packlets.
This causes the first \packlet of each \packet to be dropped, and 
thus, slows down the progression of the flood. 
In particular, \name (compliant) needs three instead of 
two \packlets per hop to progress.   
Second, the radio hardware will set the footer of the \packet, i.e., the gray-shadowed footer in Fig.~\ref{fig:packet_format}. 
To avoid this footer to collide with the footer of 
synchronously sent 
\packlets, \name (compliant) makes that all nodes stop 
sending at the same time, so that the footers of all \packets align. 

\subsection{Resilience against external interferences}
\label{subsec:channel_hopping}

As other approaches based on synchronous transmissions, \name is sensitive to external interference. 
Common devices such as microwave ovens or Wi-Fi access points can disturb communication and significantly reduce the reliability of the protocol.
Several approaches already presented in the literature show that introducing frequency diversity -- in particular channel hopping -- is an effective countermeasure against external interference~\cite{Istomin2018+Crystal2,Lim2017+Competition,Sommer2016+GlossyHopping}.
These techniques, especially sending each flood on a different frequency~\cite{Istomin2018+Crystal2}, are straightforward to integrate in \name.

\subsection{Porting \name to other radios}
\label{subsec:porting}

Whisper exploits low-level mechanisms that are, admittedly, not 
available for all radio transceivers. 
However, the design of \name is still not limited to the CC2420 radio chip and the TelosB platform. 
The CC2520~\cite{CC25202007} that is used in the WiSMote~\cite{WiSMote} nodes also provides the TXFIFO looping mode. 
Often, radio chips provide the TXFIFO looping mode functionality using a different name. 
For example, the Atmel~AT86RF215~\cite{AT86RF215} that is deployed in the OpenMote-B\footnote{\url{http://www.openmote.com/blog/openmote-b-released/}} calls it \emph{frame based continuous transmission} or \emph{continuous transmission} in the AT86RF231 chip~\cite{AT86RF231} that the M3~Open~Nodes\footnote{\url{https://www.iot-lab.info/hardware/m3/}} and A8~Open~Nodes\footnote{\url{https://www.iot-lab.info/hardware/a8/}} are equipped with. 
The Semtech~SX1211~\cite{SX1211} chip calls this functionality \emph{buffered mode} and the default mode is here called packet mode (not to be confused with the default 'buffered mode' of the CC2420). 
The CC1101~\cite{CC11012013} that is deployed in the MSP430-CCRF~\cite{Olimex2013+MSP430CCRF} nodes provides three packet length modes that can be reprogrammed during receive and transmit: fixed mode, variable mode, and infinite mode. 
Using these modes, the CC1101 supports packet lengths that are longer 
than 127 bytes. \name can use them to avoid the transmission of the 
length field of the signaling packet.

For radios that do not support the TXFIFO looping mode or similar 
options, a software-based solution like the byte-wise transmission 
power control approach from Saha and Chan~\cite{Saha2017+Many-to-one} 
-- where the length field is transmitted with the smallest possible transmit power -- as 
mentioned in Sec.~\ref{subsec:packlets_identical}, could be an interesting alternative approach that we plan to investigate in a future project.

\subsection{Using \name with large payloads}
\label{subsec:payload_size}

\begin{figure}
	\centering
	\subfloat[\emph{First hop.}]{
		\includegraphics[width=.71\linewidth]{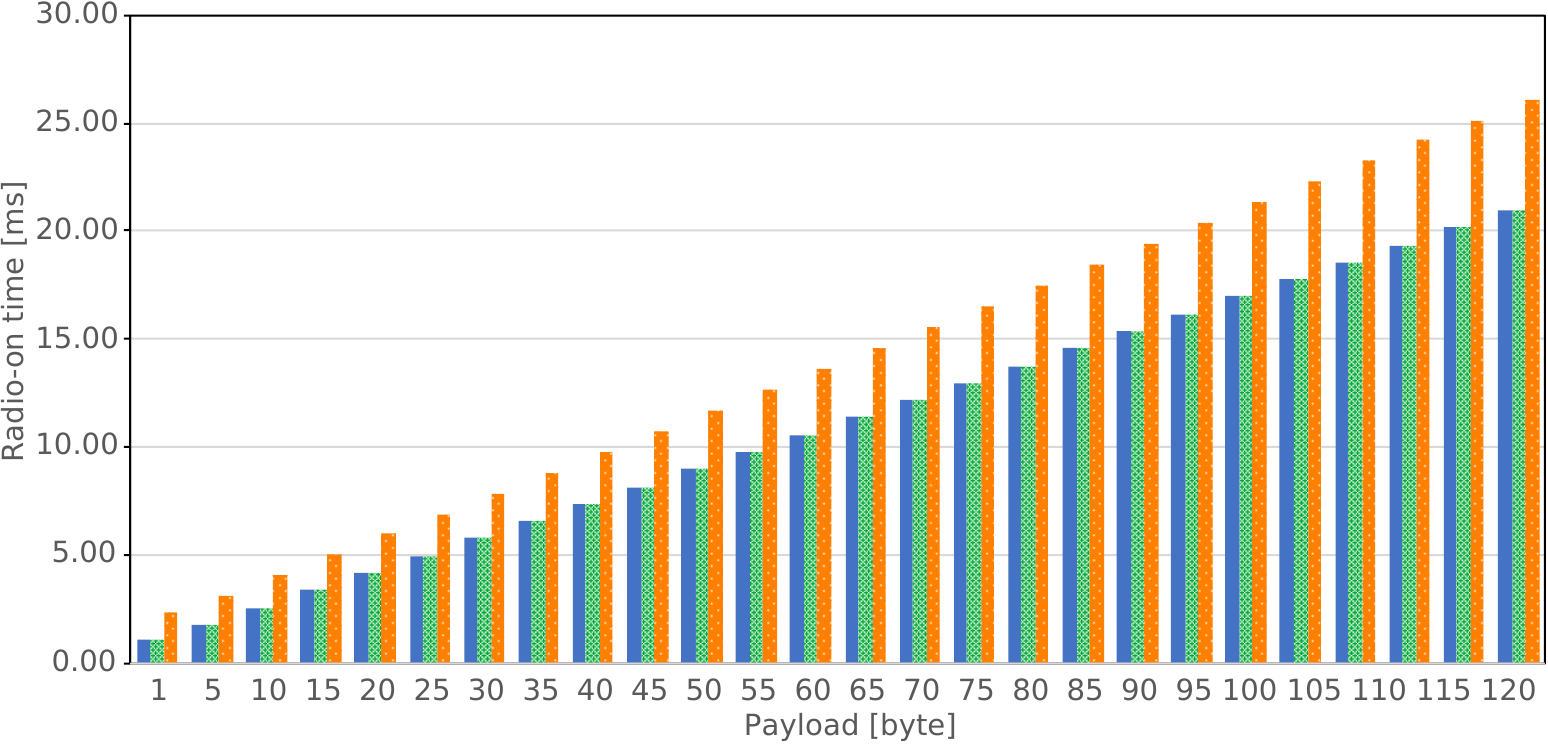}
		\label{fig:payload-size_1hop}
	}
	
	\subfloat[\emph{Third hop.}]{
		\includegraphics[width=.71\linewidth]{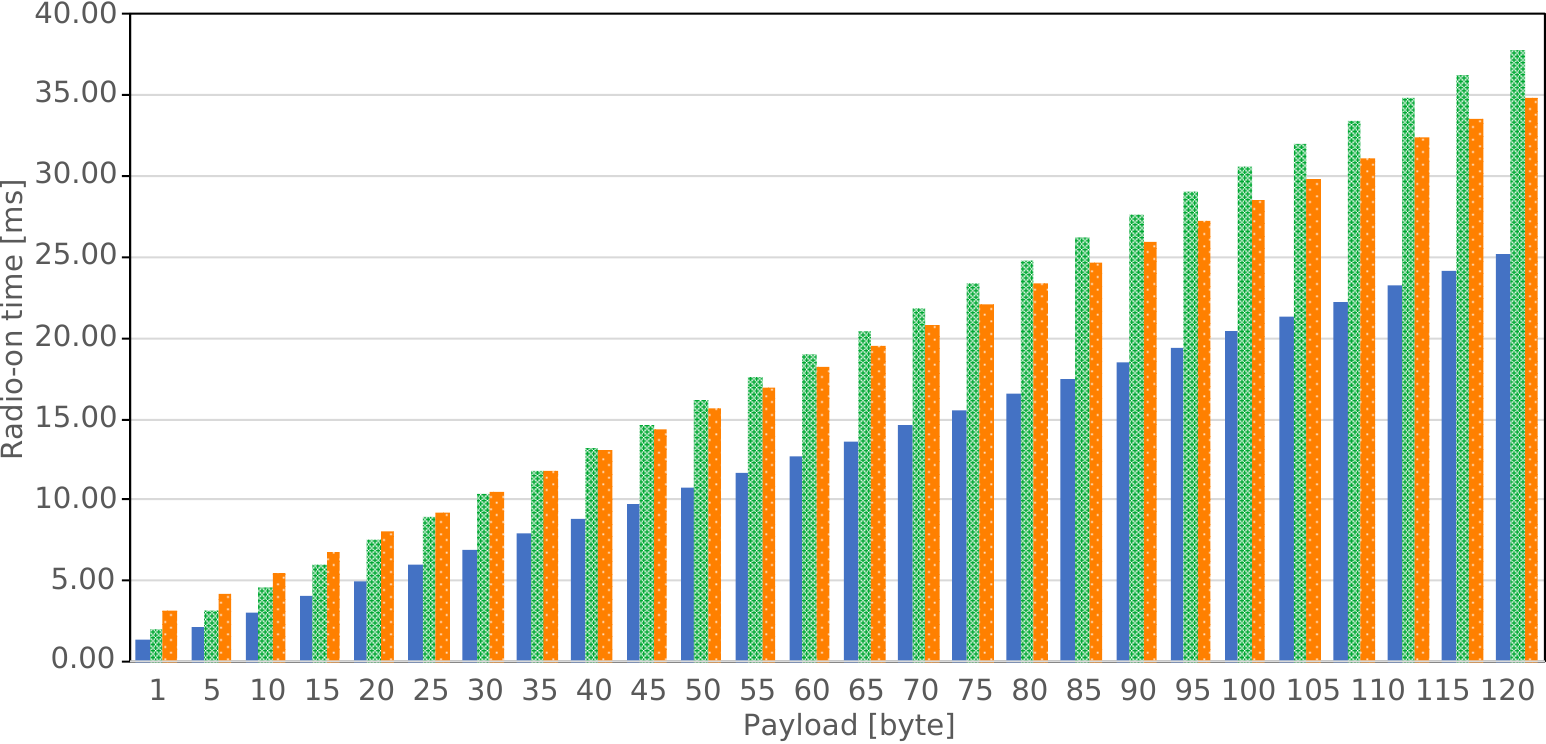}
		\label{fig:payload-size_3hop}
	}
	
	\subfloat[\emph{Sixth hop.}]{
		\includegraphics[width=.73\linewidth]{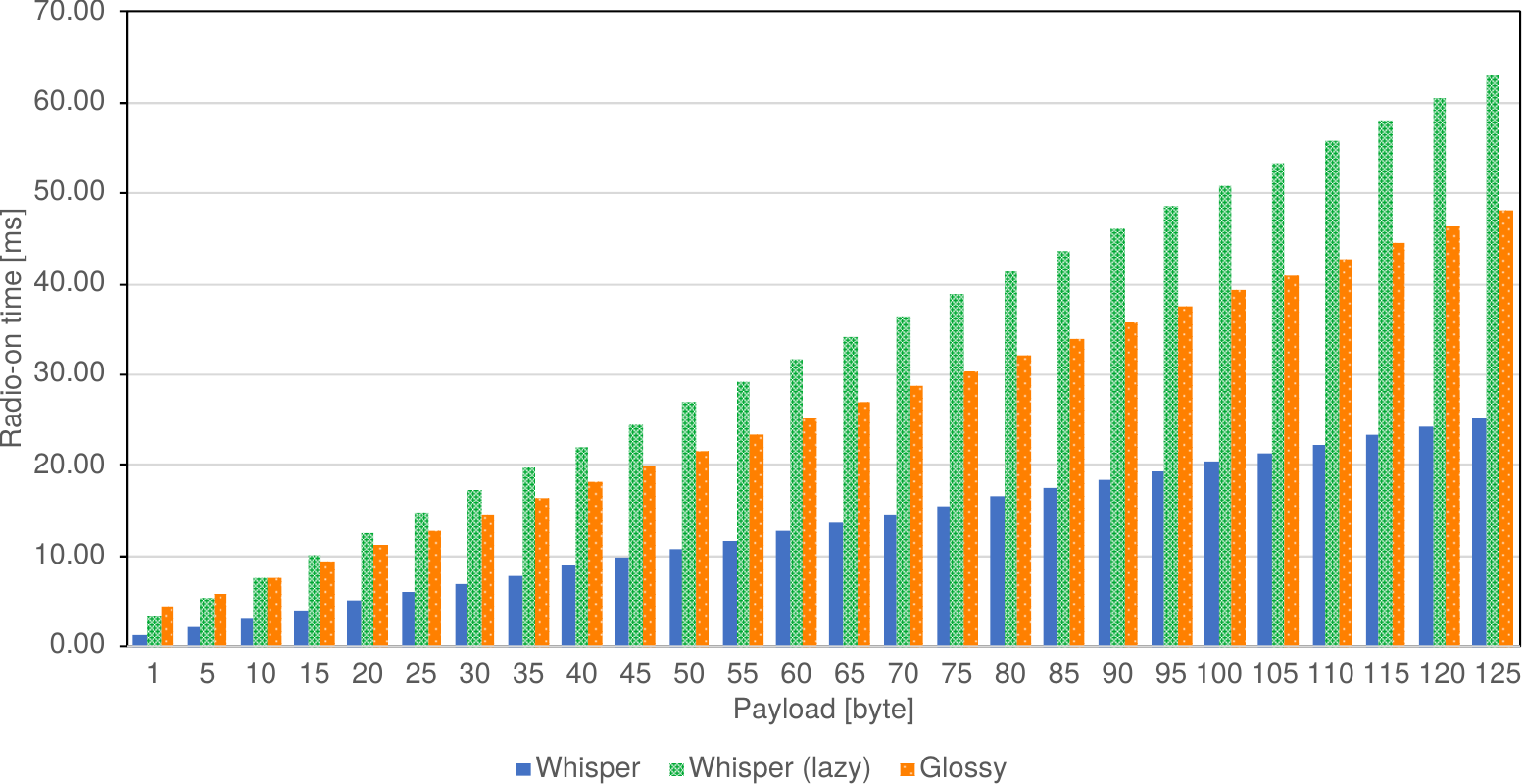}
		\label{fig:payload-size_6hop}
	}
	\caption
	{
		\emph{The theoretical radio-on time for different payloads at different hops.}
		The nodes in the first hop using \name and \name~(lazy) achieve a lower radio-on time compared to the nodes running \glossy for all payload sizes.
		The nodes using \name (lazy) in the third hop exceed the radio-on time of the nodes using \glossy at 40~bytes of payload and the nodes in the sixth hop running \name~(lazy) exceed the nodes running \glossy's radio-on time at 15~bytes payload.
		\name achieves for all three network diameters and all 
		payload sizes a lower radio-on time compared to \glossy.
	}
	\label{fig:payload-size}
\end{figure}

This paper describes \name as quick yet energy-efficient flooding primitive for small amounts of data.
However, \name also supports the propagation of large data, i.e., \packlet lengths of 127~bytes.
Using the TXFIFO looping mode, the radio returns to the beginning of the TXFIFO buffer when the radio has reached the end of the buffer. 
The MCU has, thus, to continuously fill the buffer while the radio continuously reads the buffer. 
This allows a sender to transmit \packlets and \packets of infinite length. 
However, the receivers operate in buffered mode in which the length of a \packlet is limited to 127~bytes (while the length of a \packet is still unlimited).

Even though \name also works for large data, the sampling strategies presented in this paper are designed for small data.
For example, in the direction-aware sampling strategy, the nodes wake-up one \packlet before the actually expected \packlet.
Transmitting and receiving a 127~byte \packlet takes about 4~ms. 
Thus, the nodes stay in idle listening for the entire 4~ms, which is power inefficient. 
In applications with large data, a new sampling strategy should 
be designed, e.g., by letting the nodes only sample at the 
beginning of a potential \packlet arrival instead of the entire 
\packlet duration.

We show in Fig.~\ref{fig:payload-size} the theoretical radio-on time for different payloads and at different hops for \name, \name~(lazy), and \glossy.
\glossy exceeds the radio-on time of \name and \name~(lazy) for all payload sizes in the first hop, shown in Fig.~\ref{fig:payload-size_1hop}.
However, at the third hop, \name~(lazy) has a higher radio-on time compared to \glossy and \name with a payload of 40~bytes, depicted in Fig.~\ref{fig:payload-size_3hop} and at the sixth hop at a payload of 15~bytes as shown in Fig,~\ref{fig:payload-size_6hop}.
This is because \name's progression of 2~\packlets per hop, and thus, a node in the third hop running \name~(lazy) spends a duration of 4~\packlets in idle listening.
In comparison, a node running \glossy in the same configuration only is in idle listening for 2 packets and 2-times the RX/TX turnaround.
\name using direction-aware sampling achieves a low radio-on time in all hops, showing that \name can also be used for large data.
For a more detailed comparison of the theoretical radio-on time of \name, \name~(lazy), and \glossy, we refer to Table~\ref{tab:cmp_radio-on} in Appendix A.

\section{Evaluating \name in flooding scenarios}
\label{sec:eval}

In this section, we evaluate \name in extensive testbed experiments.
In this set of experiments, we consider scenarios in which 
\name disseminates periodically small data of a few byte like 
configuration parameters.
We find that \name has a 50\% higher network lifetime compared to 
\glossy while achieving a reliability which is close to 100\%.

\subsection{Evaluation setup}
\label{sec:eval_setup}

\paragraph{Implementation}
We implemented \name for the Contiki operating system\footnote{\url{http://www.contiki-os.org/}}. 
We embedded the code base of Flock\footnote{\url{https://github.com/martinabr/flock}} into \name and reused parts of the publicly available implementation of \glossy\footnote{\url{http://sourceforge.net/p/contikiprojects/code/HEAD/tree/ethz.ch/glossy/}\label{fn:glossyurl}} in our code.

\paragraph{Metrics}
We focus on two key performance metrics: reliability and radio-on time.
We compute the \emph{per-node reliability} as the ratio of the total number of \packets successfully received by a node and the total number of \packets sent during an experiment.
We then derive the \emph{network reliability} as the average of the reliability of all nodes in the network.
The \emph{radio on-time} is the time the radio is turned on 
and active (including idle listening) during a \name (or \glossy) 
slot. 
As for the case of reliability, the radio-on time of the network is computed as the average of the radio-on time of each node.


\paragraph{Testbed}
We run our experiments on the FlockLab testbed \cite{Lim2013+FlockLab}.
FlockLab is an indoor testbed with 27 nodes deployed in an office 
building of the ETH Zurich in Switzerland.
The nodes available for our experiments are the Tmote Sky, equipped 
with the MSP430F1611 low-power microcontroller~\cite{msp430f1611} and 
the CC2420 IEEE~802.15.4 transceiver~\cite{CC24202013}.

\paragraph{\name and \glossy versions used in the evaluation}
To illustrate the performance of \name in detail, we implement different versions of the protocol.
\name is the full-fledged protocol that includes direction-aware sampling (see Sec.~\ref{subsec:direction_aware_sampling}) and exploits the TXFIFO looping mode (see Sec.~\ref{subsec:packlets_identical}).
In \name, we further use a 2-byte preamble as mentioned in Sec.~\ref{subsec:packet} and set $N_{tx}=3$. 

We also explore the performance of \name in a series of other configurations, e.g., with lazy sampling instead of direction-aware sampling, with a 4-byte instead of 2-byte preamble, as well as with different values of $N_{tx}$.
In the plots, we indicate after the name of \name the specific change with respect to the default implementation, i.e, ``\name (lazy)'' indicates a version of \name that uses lazy sampling but keeps the TXFIFO looping mode, the 2-byte preamble and $N_{tx}=3$.
Lastly, we also consider the fully IEEE 802.15.4-compliant version of 
\name described in Sec.~\ref{subsec:whisper_compliant}. 
\name (compliant) uses a 4-byte preamble, lazy sampling, $N_{tx}=14$ and does not exploit the TXFIFO looping mode.

As for \glossy, we use its publicly available code base\footref{fn:glossyurl}. 
As discussed in Sec.~\ref{subsec:packet}, we set the payload of \glossy packets to 1 byte to avoid an unfair penalization of \glossy due to its larger packet size. 
We provide experimental results obtained by running \glossy with both a 2 byte and a 4~byte~preamble.




\newcommand{\disssingle}{diss.~fixed}
\newcommand{\dissvary}{diss.~diff.}
\newcommand{\dissclose}{diss.~close}
\newcommand{\dissdistributed}{diss.~far}

{\footnotesize \begin{table}[htbp]
		\caption{\emph{Summary of scenarios description and configuration parameters.}}
		\begin{tabular}{l l l}
			\toprule
			Scenario & Label & Sender \\ 
			& & \emph{[node id in FlockLab]} \\
			\midrule
			Dissemination with fixed sender & \disssingle & 1 \\
			Dissemination with different senders & \dissvary & 10, 22, 11, 16, 23, 19, 20, 31, 26, 7 \\
			Dissemination with concurrent, close-by senders &\dissclose & 4, 2, 8, 1  \\        
			Dissemination with concurrent, far-away senders & \dissdistributed & 16, 19, 7, 1  \\
			\bottomrule
		\end{tabular}
		\addtolength{\tabcolsep}{1pt}
		\label{tab:overview_scenarios}
\end{table}}

\paragraph{Scenarios}
We run experiments in different dissemination scenarios, as 
summarized in Table~\ref{tab:overview_scenarios}, in which nodes 
flood small amounts of data into a multi-hop network.
We test both dissemination with only one sender and with different senders. 
We also consider the case in which different senders transmit 
concurrently and differentiate between concurrent senders positioned 
close-by each other or roughly evenly distributed across the network.
A summary of our evaluation results for the different 
	scenarios can be found in Table~\ref{tab:eval_scenarios}. However, in 
	the following, we describe our results in more detail.
\subsection{\name vs. \glossy}
\label{sec:eval_compare_glossy}

\begin{figure}
	\centering
	\subfloat[
	\emph{Performance during data dissemination.}
	Comparing with Glossy reduce \name and \name (lazy) the radio 
	on-time by a factor of two while achieving a reliability near 
	100\%.
	Nodes have learned their distance to the source (node~1) and efficiently turn their radio on before the flood \enquote{passes-by}.
	]{%
		\includegraphics[width=\textwidth]{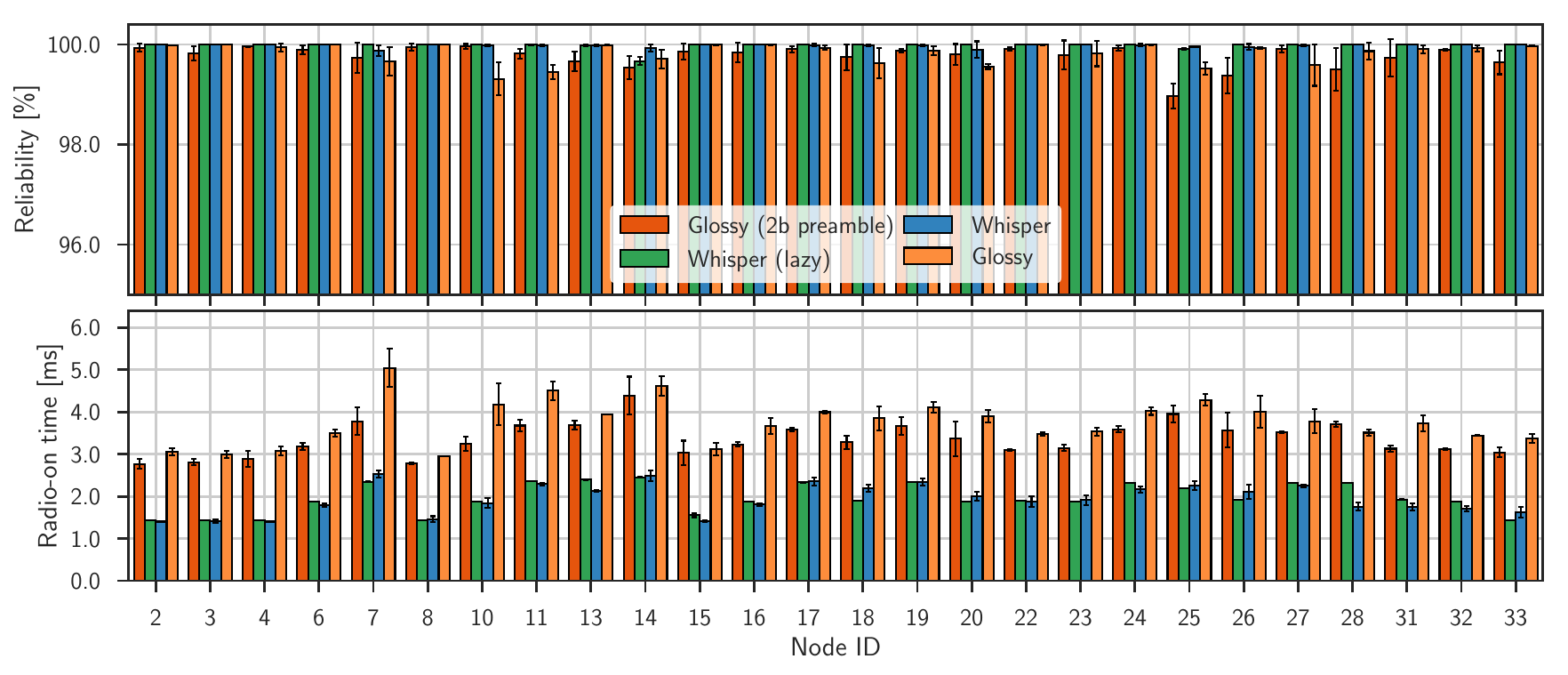}
		\label{fig:plot_summary}
	}
	
	\subfloat[
	\emph{Radio on-time when no \packet is disseminated.}
	\name achieves a low radio on-time even when no \packet is disseminated. 
	In contrast, \name (lazy) and \glossy use a fixed timeout mechanism that is set to 5~ms (marked as red line) to turn the radio off in case no \packlet has been received.
	Learning the distance to the source (node~1) allows for energy-efficient channel checks.
	]{%
		\includegraphics[width=.69\textwidth]{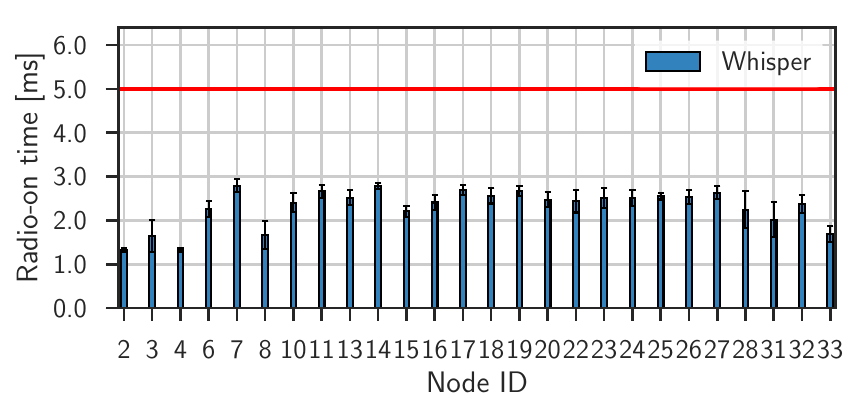}
		\label{fig:plot_summary_idle_listening}
	}
	\caption{
		\emph{Performance of \name, \name (lazy), \glossy, and \glossy (2b preamble) at 0~dBm in FlockLab in a data dissemination scenario with only a single, fixed sender (\disssingle).}
		\name outperforms \glossy in terms of energy-efficiency during data dissemination as well as when no \packet has been sent.
	}
	\label{fig:diss_single}
\end{figure}

We first compare the performance of \name, \name (lazy) as well as \glossy and \glossy (2b preamble) in a dissemination scenario with a single, fixed sender (\disssingle).
This scenario corresponds to, e.g., a controller that needs to signal the nodes to stay awake for an unscheduled software update or to disseminate some configuration parameters to all nodes in the network.
We find that nodes using \name achieve, with respect to \glossy, a comparable or higher reliability and a significantly smaller radio-on time, both with and without data traffic.

\paragraph{Experiments}
We run \name, \name (lazy), \enquote{standard} \glossy and \glossy (2b preamble) in the following configuration.
We select the node with identifier 1 as the sender. 
This node is located on the outer edge of the FlockLab testbed, which allows us to obtain a large network diameter.
To vary the topology and in particular the number of hops between the sender and the farthest receivers, we use two different transmit powers: -10~dBm and 0~dBm.
This results in a network diameter of 3 to 4 and 5 to 6 hops, respectively.
Each experiment consists of 10'000 floods and we repeat each experiment 3 times.
For \name, we further measure the radio-on time over 5'000 slots during which the sender sends no \packet.
The collected per-node data is averaged over the three independent 
runs and the standard deviation is plotted as error bars in 
the figures.


\paragraph{Results}
Fig.~\ref{fig:plot_summary} details -- for the case in which the sender disseminates a \packet in each slot -- the per-node reliability in the upper plot and the per-node radio-on time in the lower plot.
While the reliability of \name and \name (lazy) is comparable or 
slightly higher than that of \glossy and \glossy (2b preamble), the 
radio-on time is significantly lower -- on average half that of 
\glossy\xspace -- for \name and \name (lazy).
Assuming that the protocols are scheduled every second, \name has a 
50\% higher network lifetime than \glossy\footnote{
	The average radio-on time is 1.9~ms and 3.7~ms for \name and \glossy, 
	respectively.
	We assume a battery capacity of 2000~mAh and consider only the energy-consumption during communication (i.e., 20~mA).
	The resulting network lifetime is 2193~days and 1126~days for \name and \glossy, respectively.
	Thus, the network lifetime of \name is 50\% higher than the one of 
	\glossy.
}.
Fig.~\ref{fig:plot_summary} also shows that \name and \name~(lazy) achieve a similar radio-on time. 
However, with increasing network diameter, the radio-on time of 
\name~(lazy) significantly increases compared to \name, as shown in 
Table~\ref{tab:eval_scenarios} at -10~dBm.

Fig.~\ref{fig:plot_summary_idle_listening} shows the per-node radio-on time of \name when no \packet is sent.
The bold, (red) line at 5~ms corresponds to the radio-on time of approaches like \name~(lazy) or \glossy that  -- in case of the absence of communication -- keep nodes in idle listening for the entire slot. 
\name can save radio-on time by a factor of 2 in this case thanks to 
the use of direction-aware sampling, which makes nodes switch their 
radio off at most when the expected reception time of the \packlet 
with counter $c = c_{max} + N_{tx} + 1$  has elapsed. 
This characteristic of \name is particularly relevant when nodes must frequently switch on their radios to limit delays in relaying data traffic -- yet often no packet is flooded, like in the data prediction scenario of Crystal~\cite{Istomin2016+Crystal}. 


{\footnotesize \begin{table}
		\caption{
			\emph{Summary of evaluation results.}
			Whisper and Whisper (lazy) outperform \glossy in terms of reliability and radio-on time in various scenarios.
			Nodes using Whisper (lazy) and Glossy use a timeout mechanism to turn the radio off in case they have not intercepted a \packlet/packet within a given time.
			In this evaluation the timeout is set to 5~ms.
		}
		\begin{tabular}{p{9ex} l *{4}{c}}
			\toprule
			Protocol & Scenario & Tx power & Reliability & Radio on  & Radio on   \\
			& & & & w/ signaling & w/o signaling \\
			& & [dBm] & [\%] & [ms] & [ms]\\
			\midrule
			\multirow{2}{*}{\celll{h}{9ex}{\name}} & \multirow{2}{*}{\disssingle} 
			& -10 & 99.980 & 2.055 & 2.546 \\
			&  & 0   & 99.980 & 1.936 & 2.474 \\ 
			\midrule   
			
			\multirow{8}{*}{\celll{h}{9ex}{\name (lazy)}} & \multirow{2}{*}{\disssingle} 
			& -10 & 99.817 & 2.477 & 5.0 \\
			& & 0   & 99.983 & 1.962 & 5.0 \\ 
			& \multirow{2}{*}{\dissvary} 
			& -10 & 99.932 & 2.175 & 5.0 \\
			&  & 0   & 99.986 & 1.865 & 5.0 \\
			& \multirow{2}{*}{\dissclose} 
			& -10 & 99.887 & 2.438 & 5.0 \\
			&  & 0   & 99.952 & 2.129 & 5.0 \\
			& \multirow{2}{*}{\dissdistributed} 
			& -10 & 99.786 & 1.626 & 5.0 \\
			&  & 0   & 99.965 & 1.540 & 5.0 \\
			\midrule

			\multirow{2}{*}{\celll{h}{9ex}{\glossy}} & \multirow{2}{*}{\disssingle} 
			& -10 & 99.738 & 4.253 & 5.0 \\
			&  & 0   & 99.828 & 3.756 & 5.0 \\ \midrule
			
			\multirow{8}{*}{\celll{h}{9ex}{\glossy (2b preamble)}} & \multirow{2}{*}{\disssingle} 
			& -10 & 99.616 & 3.914 & 5.0 \\
			&  & 0   & 99.767 & 3.356 & 5.0 \\ 
			& \multirow{2}{*}{\dissvary} 
			& -10 & 98.369 & 3.805 & 5.0 \\
			&  & 0   & 98.963 & 3.351 & 5.0 \\
			& \multirow{2}{*}{\dissclose} 
			& -10 & 99.350 & 4.071 & 5.0 \\
			&  & 0   & 99.024 & 3.932 & 5.0 \\
			& \multirow{2}{*}{\dissdistributed} 
			& -10 & 98.881 & 3.680 & 5.0 \\
			&  & 0   & 98.559 & 3.721 & 5.0 \\
			\bottomrule
		\end{tabular}
		\addtolength{\tabcolsep}{1pt}
		\label{tab:eval_scenarios}
\end{table}}

\subsection{\name in dissemination scenarios}
\label{sec:eval_dissemination}


\begin{figure}
	\centering
	\includegraphics[width=0.65\textwidth]{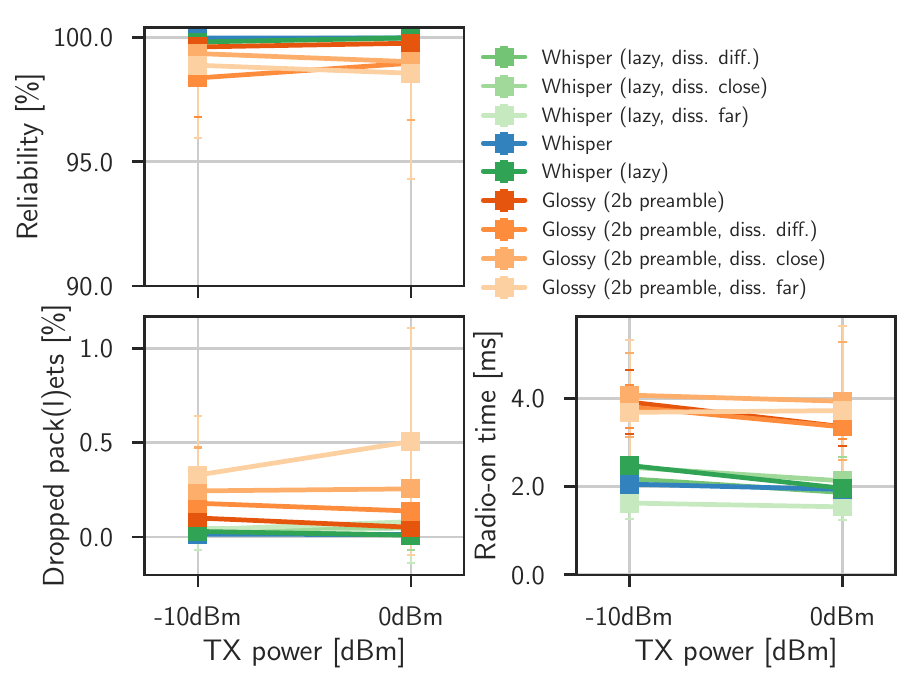}
	\caption{
		\emph{Comparison in dissemination scenarios.}
		\name and \name (lazy) achieve in all scenarios a two-fold lower radio-on time compared to \glossy. 
		At the same time, they achieve a higher reliability.}
	\label{fig:eval_dissemination}
\end{figure}


To consider the case in which different nodes must disseminate data -- possibly even concurrently -- we evaluate the performance of \name, \name (lazy) and \glossy in the three scenarios \textit{\dissvary}, \textit{\dissclose}, and \textit{\dissdistributed}\xspace (see Table~\ref{tab:eval_scenarios}).
We find that contention for the same slot causes less packet 
collisions, and thus, results in higher reliability in 
\name and \name (lazy) compared to \glossy.

\paragraph{Experiments}
We run \name, \name (lazy) and \glossy consecutively with transmit powers -10~dbm and 0~dBm. 
In the \textit{\dissvary}\xspace scenario, each sender --  shown in 
Table~\ref{tab:overview_scenarios} -- consecutively transmits 1'000 
\packets before handing over to the next sender.
The senders in the \textit{\dissclose}, and 
\textit{\dissdistributed}\xspace scenarios concurrently transmit 
\packets in every \name slot.
We execute 10'000 floods in each experiment (i.e., for each protocol) 
and we run each experiment 3~times.

\paragraph{Results}
Fig.~\ref{fig:eval_dissemination} shows the network reliability (upper plot, left), radio-on time (lower plot, right) and the percentage of dropped \packlets/packets per \name/\glossy slot. 
In all the considered scenarios, \name and \name (lazy) achieve a higher reliability and a lower radio-on time than~\glossy.

The difference in performance is more evident in scenarios with concurrent senders, i.e., \textit{\dissclose}\xspace and \textit{\dissdistributed}.
The reason is that interference due to concurrent floods has a stronger impact in \glossy than in \name.
More precisely, floods from different senders overlap with a slightly different temporal displacement caused by (i) senders not being synchronized within sub-microseconds and (ii) as stated in ~\cite{Mohammad2017+Syncast} \emph{\enquote{a combination of software, hardware, and signal propagation delays}} caused by an increasing number of concurrent transmitters.
While (i) affects both \name and \glossy to an equal extent, (ii) intensifies for each gap between consecutive transmissions, resulting in a stronger impact on \glossy compared to \name.
The consequence is that nodes using \glossy drop more packets on average, e.g., 0.5\% in \textit{\dissdistributed} resulting in 1\% lower reliability compared to \name~(lazy).

\subsection{Impact of low-level mechanisms}
\label{sec:eval_microbenchmark}

We now investigate the impact of the individual low-level mechanisms used in \name. 
We thereby consider a dissemination scenario with a single, fixed initiator (\textit{\disssingle}). 

\begin{figure}
	\centering
	\subfloat[
	\emph{Impact of preamble length.}
	A 4~byte preamble increases the overall reliability by 0.1\% while increasing the radio-on time by 20\% and 10\% for \name and \glossy, respectively, compared to a 2~byte preamble.
	]{%
		\includegraphics[width=.48\textwidth]{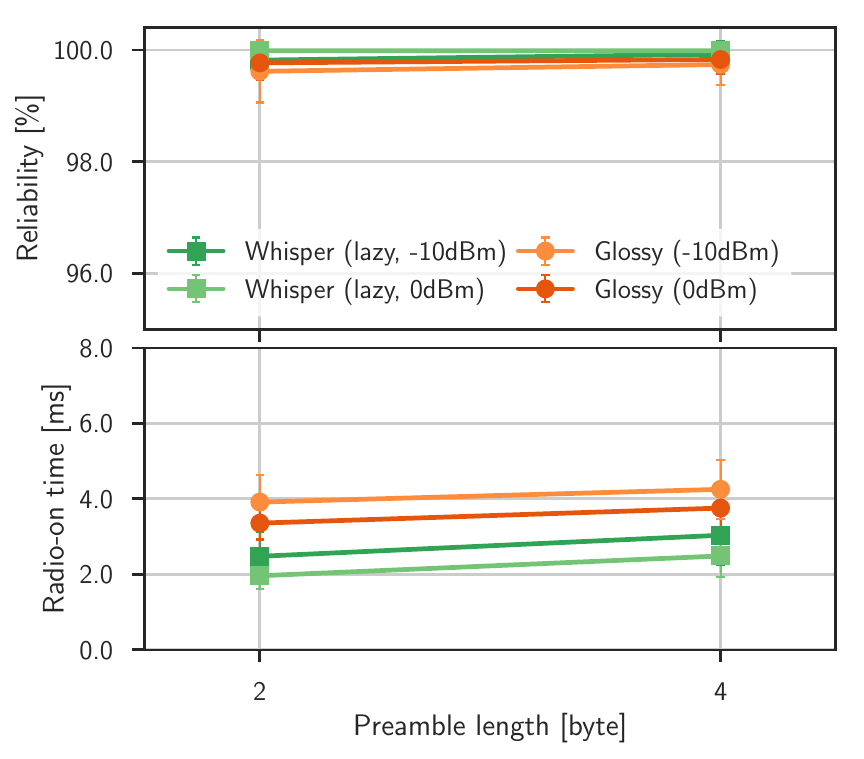}
		\label{fig:eval_preamble}
	}
	\quad
	\subfloat[
	\emph{Impact of number of \packlet/packet transmissions.}
	Glossy's radio-on time increases stronger with $N_{tx}$ compared to \name's (lazy, 4b preamble).
	The reason is the additional packet reception as well as the RX/TX turnaround.
	]{%
		\includegraphics[width=.48\textwidth]{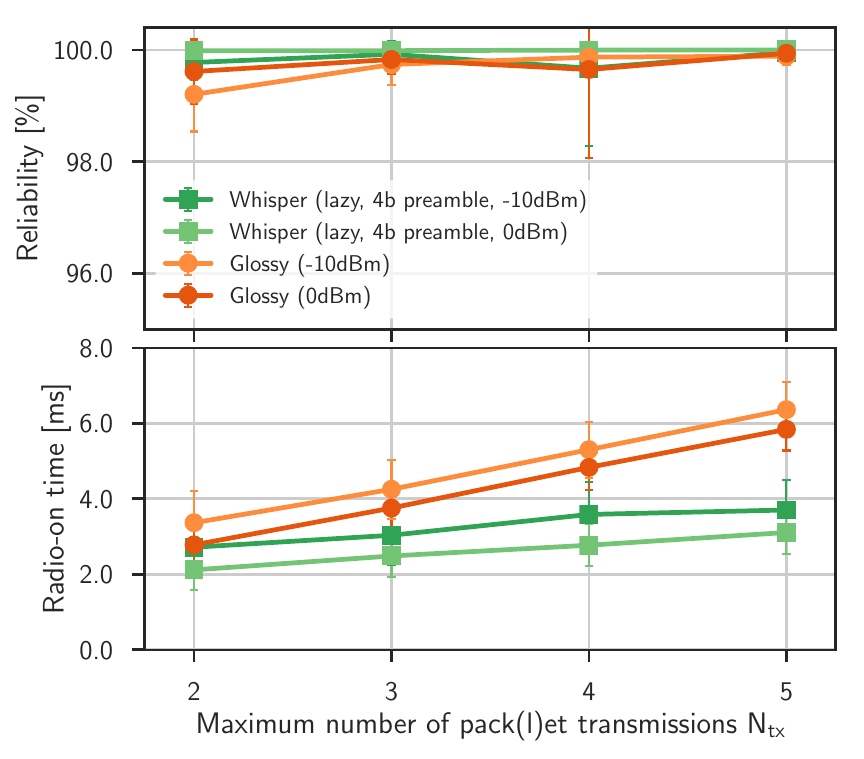}
		\label{fig:eval_ntx}
	}
	\quad
	\subfloat[
	\emph{Impact of length field.}
	Colliding lenght fields in \name (compliant) have a great impact on the progression speed of the flood.
	]{%
		\includegraphics[width=.7\textwidth]{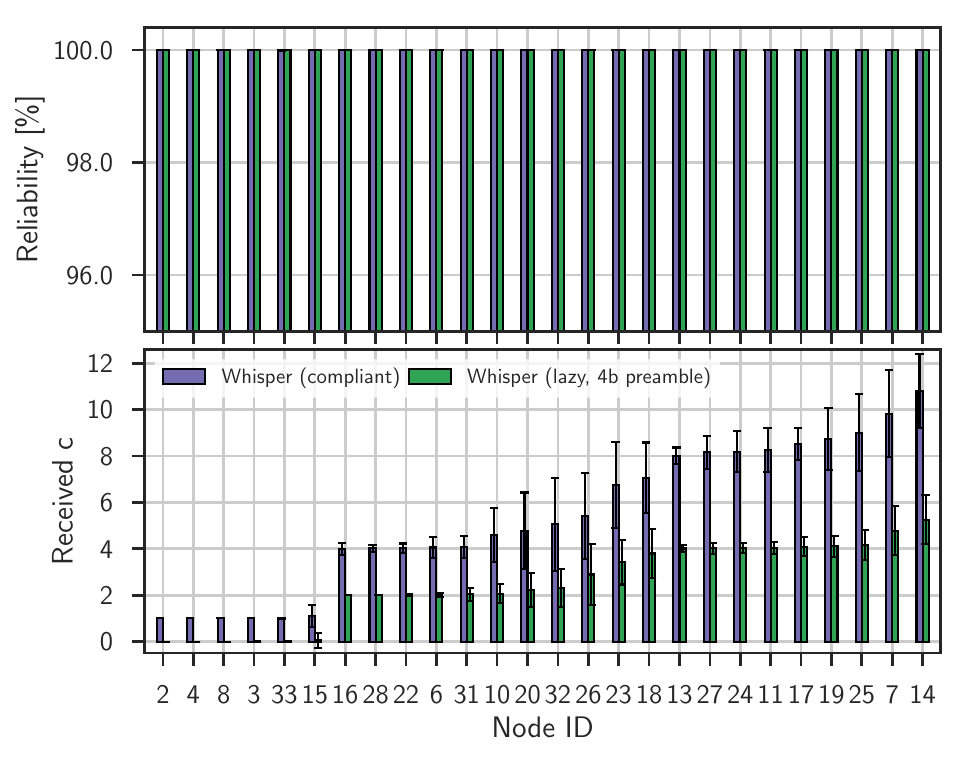}
		\label{fig:eval_length}
	}
	\caption{\emph{Impact of low-level mechanisms.}}\label{fig:animals}
\end{figure}

\subsubsection{Impact of preamble length}
\label{sec:eval_microbenchmark_pramble}

We start by analyzing the effect of the preamble length on the performance of \name (lazy) and \glossy.
We find that a 2~byte preamble significantly reduces the radio-on time for both protocols while causing a neglibile loss in terms of reliability.


\paragraph{Experiments}
We run \name (lazy) and \glossy in the \disssingle\xspace 
	scenario using $N_{tx}=3$, preamble length of both 2~bytes and 
4~bytes, and transmit powers of -10~dBm and 0~dBm.
We execute 10'000 \name/\glossy network floods for each protocol and collect data from 3 independent runs.

\paragraph{Results}
Fig.~\ref{fig:eval_preamble} shows the network reliability in the 
upper plot and the achieved radio-on time in the lower plot.
One can observe a slight increase in reliability with the 4~byte preamble compared to the 2~byte preamble. 
Comparing the gray-shadowed results corresponding to $N_{tx}=3$ in Table~\ref{tab:eval_coverage}, the network reliability with a 2 byte preamble drops about 0.1\% for all protocols and transmit powers, which corresponds to the loss of 10 packets out of 10,000, on average.
The radio-on time, however, increases with the longer preamble by 10\% and 20\% for \name (lazy) and \glossy, respectively, as shown in Fig.~\ref{fig:eval_preamble}.
To an almost negligible decrease of reliability, thus, 
corresponds a significant improvement in terms of radio-on time. 
This can be explained considering that the preamble and SFD byte are used by receivers to achieve symbol synchronization and to adjust for frequency offsets~\cite{CC24202013}.
The length of the preamble, however, only affects transmissions.
The receiver starts intercepting a packet as soon as it has found a single preamble byte followed by the SFD. 
Transmitting a longer preamble is useful to increase the signal-to-noise ratio, and thus, to help the receiver in detecting the preamble and SFD bytes.
An increase of the preamble length from 2 to 4 bytes leads, however, to almost negligible improvements, as illustrated above. 

\subsubsection{Impact of the number of transmissions $N_{tx}$}
\label{sec:eval_microbenchmark_ntx}

We now discuss how different values of $N_{tx}$ affect the performances of both \name and \glossy.
We find that both protocols achieve a similar reliability. 
However, \name (lazy, 4b preamble) has a smaller radio-on time than \glossy and with every $N_{tx}$, the effect on the radio-on time increases stronger in \glossy compared to \name (lazy, 4b preamble).

\paragraph{Experiments}
We run \name (lazy, 4b preamble) and set \mbox{$N_{tx} = \{2,3,4,5\}$}.
We use transmit powers -10~dBm and 0~dBm and execute 10'000 \name/\glossy network floods for each protocol and collect data from 3 independent runs.
We further run \enquote{standard} \glossy with a preamble length of 4~byte in the same configuration. 

\paragraph{Results}
The upper plot of Fig.~\ref{fig:eval_ntx} shows that for different values of $N_{tx}$ \name (lazy, 4b preamble) and \glossy achieve a comparable reliability.
Apart from minor fluctuations, the reliability increases as $N_{tx}$ increases, as expected. 
The lower plot in Fig.~\ref{fig:eval_ntx} shows that \name outperforms \glossy in terms of radio-on time even with lazy sampling and 4 byte preamble.
More precisely, Table~\ref{tab:eval_radio} shows that in \name (lazy, 4b preamble) increasing $N_{tx}$ by 1 causes an increase of the radio-on time of roughly 288~$\mu$s -- which corresponds to $T_{packlet}$ for a \packlet with a 4~byte preamble and a 1-byte payload.
In \glossy the radio on-time increases for each $N_{tx}$ by the duration of one received and one transmitted packet \`{a} 288~$\mu$s, the RX/TX turnaround time with 192~$\mu$s and 23~$\mu$s for the software delay.
As a consequence, the increase in radio-on time with increasing $N_{tx}$ is more prominent in \glossy than in \name.

\subsubsection{Impact of collisions due to different length fields}
\label{sec:eval_microbenchmark_length}

Lastly, we compare \name (compliant) with \name (lazy, 4b preamble, 14~\packlets).
We find that collisions caused by the different length fields in \name~(compliant) have a significant, negative influence on the speed at which a flood can progress. 

\paragraph{Experiments}
We run \name (compliant) with 14~\packlets, which results in a \packet of 122 bytes.
We further run \name (lazy, 4b preamble, 14~\packlets) and make all nodes stop transmitting their \packets simultaneously.
The \packets in \name (compliant) and in \name (lazy, 4b preamble, 14~\packlets) differ only in the length field of the first \packlet. This is the length of the \packet in \name (compliant) and the length of a \mbox{\packlet in the latter}.



\paragraph{Results}
Fig.~\ref{fig:eval_length} shows the per-node reliability on the upper plot and the received counter $c$ on the lower plot.
We find that each node achieves a reliability of 100\% for both protocols. 
This is consistent with the results discussed in the previous Sec.~\ref{sec:eval_microbenchmark_ntx}, where we found that the reliability increases with each additionally transmitted \packlet.
This is also the case when nodes simultaneously stop sending instead of ceasing after $N_{tx}$ transmissions.


The lower plot of Fig.~\ref{fig:eval_length} reveals that nodes using \name (compliant) receive higher counter values compared to \name (lazy, 4b preamble, 14~\packlets).
This is what causes a slower progression of the flood and is not unexpected given that in \name (compliant) 
(i) nodes drop \packlets whose length field is not set correctly, and
(ii) the \packlets are exposed to collisions due to the different length fields.
More precisely, the \packlet with $c=0$ is dropped by the nodes in the first hop (nodes with identifiers 2 to 15 in Flocklab), because the length field is not set to the length of a \packlet but to the length of the \packet.
The nodes in the first hop receive \packlet $c=1$ and consequently miss $c=2$ due to the RX/TX turnaround.
They transmit \packlet $c=3$, which collides with the \packlet of the sender.
Thus, nodes on the second hop (identifiers 16 to 31) successfully receive \packlet $c=4$.
This procedure continues until the last hop (node with identifier 14) receives \packlet $c=11$.
In comparison, the same node receives $c=5$ with \name (lazy, 4b preamble, 14~\packlets).
This shows that the flood progresses faster with \name (lazy, 4b 
preamble, 14~\packlets), and thus, requires less \packlets to be sent 
in total, which reduces the radio-on time. 

{\footnotesize \begin{table}
		\caption{\emph{Summary of low-level mechanisms.}}
		\label{tab:low-level}
		\subfloat[\emph{Reliability.}]{
			\addtolength{\tabcolsep}{-3pt}
			\begin{tabular}{*{8}{c}}
				\toprule
				& & \multicolumn{4}{c}{4b preamble} & & 2b preamble \\
				\cmidrule{3-6} \cmidrule{8-8}
				Protocol & Tx power & $N_{tx}=2$ & $=3$ & $=4$ & $=5$ & & $N_{tx}=3$ \\ 
				& [dBm] & \multicolumn{4}{c}{[\%]} & & [\%] \\
				\midrule
				\name (lazy)  & -10 & 99.774 & \cellcolor{lightgray} 99.921 & 99.672 & 99.953 & & \cellcolor{lightgray} 99.817\\
				\name (lazy)  &  0  & 99.985 & \cellcolor{lightgray} 99.986 & 99.996 & 99.998 & & \cellcolor{lightgray} 99.983\\
				\glossy & -10 & 99.204 & \cellcolor{lightgray} 99.738 & 99.870 & 99.888 & & \cellcolor{lightgray} 99.616\\
				\glossy &  0  & 99.613 & \cellcolor{lightgray} 99.828 & 99.649 & 99.939 & & \cellcolor{lightgray} 99.767\\
				\bottomrule
			\end{tabular}
			\addtolength{\tabcolsep}{1pt}
			\label{tab:eval_coverage}
		}
		
		\subfloat[\emph{Radio-on time.}]{
			\addtolength{\tabcolsep}{-3pt}
			\begin{tabular}{*{8}{c}}
				\toprule
				& & \multicolumn{4}{c}{4 byte preamble} & \phantom{a} & 2 byte preamble \\
				\cmidrule{3-6} \cmidrule{8-8}
				Protocol & Tx power & $N_{tx}=2$ & $=3$ & $=4$ & $=5$ & & $N_{tx}=3$ \\ 
				& [dBm] & \multicolumn{4}{c}{[ms]} & & [ms] \\
				\midrule
				\name (lazy)  & -10 & 2.718 & \cellcolor{lightgray} 3.036 & 3.587 & 3.705 && \cellcolor{lightgray}2.477 \\
				\name (lazy)  &  0  & 2.118 & \cellcolor{lightgray} 2.487 & 2.772 & 3.109 && \cellcolor{lightgray}1.962 \\
				\glossy & -10 & 3.367 & \cellcolor{lightgray} 4.253 & 5.303 & 6.363 && \cellcolor{lightgray}3.914 \\
				\glossy &  0  & 2.781 & \cellcolor{lightgray} 3.756 & 4.834 & 5.841 && \cellcolor{lightgray}3.356 \\
				\bottomrule
			\end{tabular}
			\addtolength{\tabcolsep}{1pt}
			\label{tab:eval_radio}
		}
		
\end{table}}
\section{Evaluating \name as wake-up primitive within Crystal}
\label{sec:eval_crystal}

\begin{figure}
	\centering
	\subfloat[\emph{Crystal.}]{
		\includegraphics[width=\linewidth]{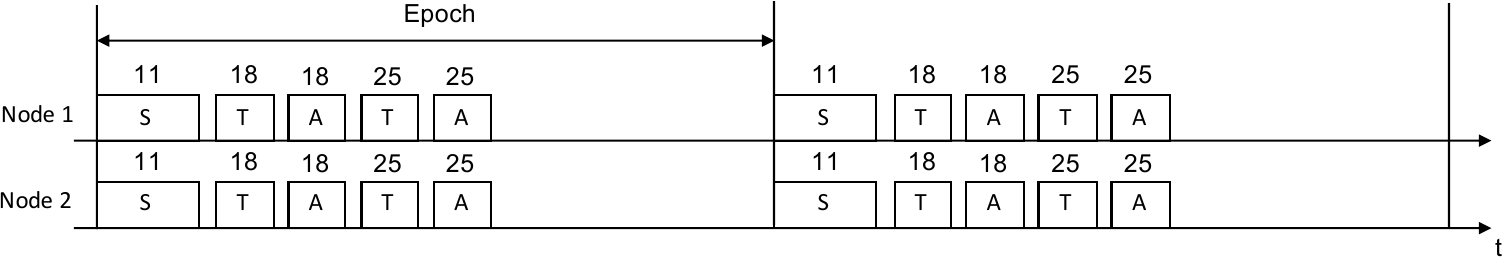}
		\label{fig:crystal}
	}
	
	\subfloat[\emph{Crystal /w \name.}]{
		\includegraphics[width=\linewidth]{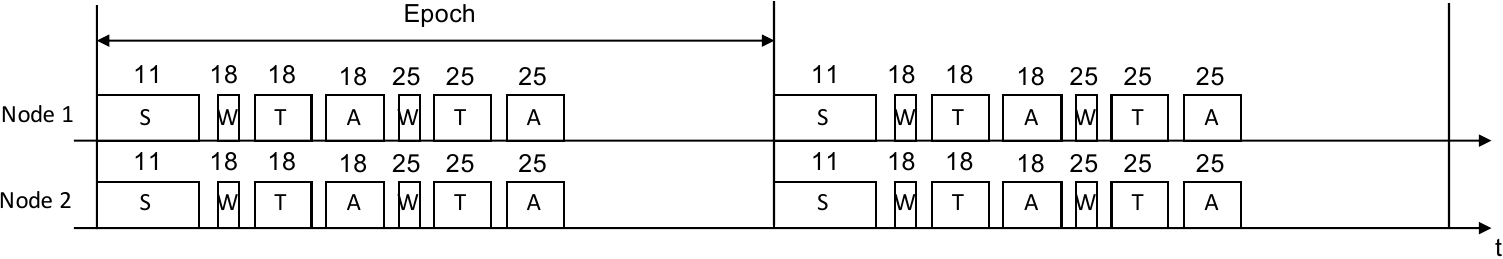}
		\label{fig:crystal_whisper}
	}
	\caption
	{
		\emph{A Crystal epoch with and without \name.}
		The (W-), T- and A-slot alternate until the sink node has 
		disseminated a negative-acknowledgment twice during 
		the A-slot.
		The numbers above the slots denote the channel used for communication.
		The illustrations are based on Figure~5 
		in~\cite{Istomin2018+Crystal2}.
	}
	\label{fig:crystal-and-whisper}
\end{figure}

In the following, we evaluate \name used as wake-up primitive in 
Crystal~\cite{Istomin2016+Crystal, Istomin2018+Crystal2}, a recently 
proposed data collection protocol based on \glossy that targets data 
prediction scenarios.
We show the performance of Crystal and Crystal with integrated \name 
and find that \name increases the network lifetime by a factor of 2.3 
compared to ``standard'' Crystal.

\paragraph{Crystal and Crystal with \name}

In the following, we briefly describe the operation of Crystal and 
afterwards how \name is incorporated in Crystal.

Crystal uses \glossy-floods to let nodes transmit data to a sink node.
When the sink node has received a data packet, it acknowledges its reception.
Otherwise, it disseminates a negative-acknowledgment.
The transmit~(T) and acknowledgment~(A) slots alternate until the sink has disseminated a negative-acknowledgment twice.
After the second negative-acknowledgment, the nodes turn their radio off.
Crystal runs periodically and organizes its operations in epochs.
Fig.~\ref{fig:crystal} shows an example of a Crystal epoch.
Each epoch starts with a synchronization~(S) slot in which the nodes re-synchronize to the sink node.
After the synchronization slot follow the T- and A-slots as described above.
To increase resilience against interference, Crystal integrates a channel hopping mechanism.
The numbers above the communication slots in Fig.~\ref{fig:crystal} indicate the channel at which the nodes communicate.
In particular, each TA-pair uses the same channel.

The sink node always disseminates packets in the S- and A-slot.
However, there is infrequent communication in the T-slot.
For example, in Crystal's temperature prediction scenario, there is no data dissemination in over 80\% of the T-slots.
However, the duration of the T-slot must be sufficiently long to support the packet size that is required by the application.
As a consequence, the nodes consume unnecessary energy.
To reduce the idle listening time when no data communication is needed, and thus, to reduce the energy consumption, \name can be integrated into Crystal as wake-up primitive.
Fig.~\ref{fig:crystal_whisper} shows Crystal with \name.
\name~(W) runs before each T-slot.
A node that has data to share, transmits a \packet in the \name slot.
The \packlets in the \packet only carry the counter $c$ as payload, 
and thus, the data field has a length of zero.
Nodes that intercept a \packlet turn their radio on in the T-slot, otherwise they keep the radio turned off until the following A-slot.
In this setting, \name relies on the synchronization and channel hopping mechanisms from Crystal.
More precisely, Crystal initiates the start of the \name slot and 
also provides the communication channel, which is the same as used 
for the TA-slots.

\paragraph{Experiments}

\begin{table}
	\centering
	\caption{
		\emph{Crystal's configuration parameters.}
	}
	{\footnotesize \begin{tabular}{c *{7}{c}}
			\toprule
			\emph{Tx power} & \emph{$N_S$} & $N_T$ & $N_A$ && $W_S$ & $W_T$ & $W_A$ \\
			\emph{[dBm]}    & \emph{[\#]}  & \emph{[\#]}  & \emph{[\#]} && \emph{[ms]}  & \emph{[ms]} & \emph{[ms]} \\
			\midrule
			0 & 3 & 2 & 3 && 10 & 9 & 7 \\
			-10 & 4 & 3 & 4 && 14 & 14 & 12 \\
			\bottomrule
	\end{tabular}}
	\label{tab:crystal_setup}
\end{table}

In this experiment, we measure the \emph{duty cycle} and the 
achieved \emph{reliability}.
In particular, the duty cycle is the averaged per-node duty cycle that is the ratio of the sum of the radio-on time for all slots during an epoch and the duration of the epoch.
The reliability indicates how many nodes that have transmitted an update in the current epoch also received an acknowledgment from the sink.
Thus, the measured reliability includes the packet reception 
rates of the T-slot, the A-slot, and when used, also the 
\name~(W)-slot.

Since Crystal is a collection protocol, we run \name using a 
``reversed'' direction-aware sampling strategy, henceforth called 
\name~(coll.). 
Thereby, the nodes must know their distance in hops to the sink, which they can learn through a short initialization phase. 
The nodes can then derive their position in the collection tree by subtracting their distance in hops to the root from the network diameter.
We use Crystal's bootstrapping period -- which lasts 10~epochs -- as initialization phase for \name to let the nodes learn their position in the network tree.
We run Crystal with an epoch of 1.5~s and a data payload of 20~bytes.
We use the default number of retransmissions~$N$ for the different slots and transmit powers, given in~\cite{Istomin2016+Crystal} and summarized in Table~\ref{tab:crystal_setup}.
As also shown in the table, we further use the default 
duration~$W$ for the S- and A-slots.
However, since we increase the payload, we also adjust the 
duration for the T-slot.
We run Crystal in FlockLab, with node 1 as sink, and with updates $u=0$, $u=2$, and $u=5$.
An update $u=X$ implies that among the 27 nodes in FlockLab, there are X events (or data packets) to share during a specific epoch.
Crystal with $u=0$ indicates that no message is transmitted in the T-slots, and thus, in this case we measure the idle listening time of Crystal.
More precisely, $u=0$ indicates the minimal power consumption, and we, thus, use it as baseline for other update configurations.
After running Crystal 3 times for 60~minutes, we repeat the 
experiments with Crystal using \name (coll.).

\paragraph{Results}

\begin{figure}[h!]
	\centering
	\includegraphics[width=.6\linewidth]{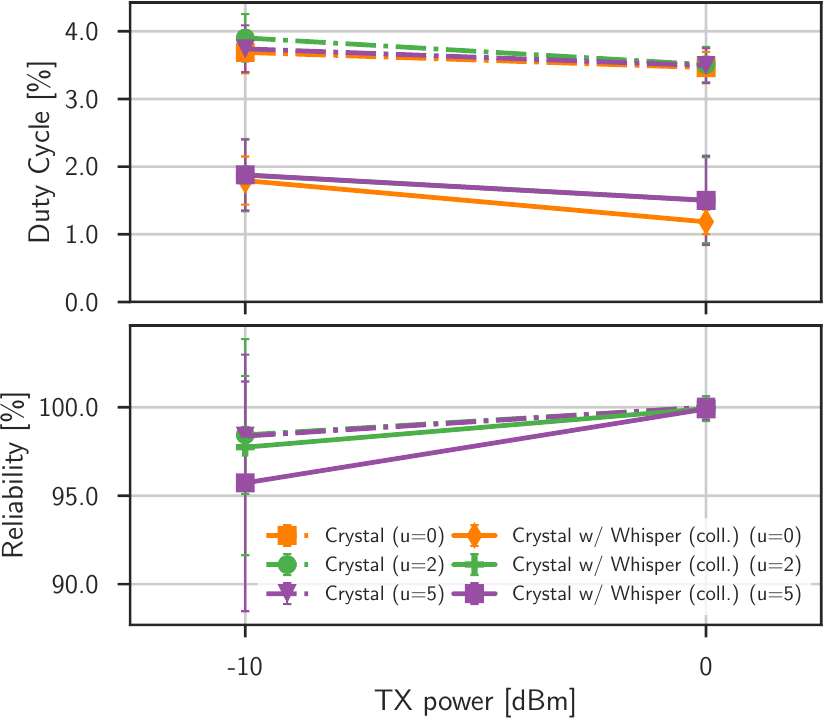}
	\caption[Impact of \name on Crystal]{
		\emph{Impact of \name on Crystal.}
		\name halves the duty cycle of Crystal. 
		However, a missed \packlet in \name also impacts the 
		reliability of Crystal.
	}
	\label{fig:plot_crystal-whisper}
\end{figure}

Fig.~\ref{fig:plot_crystal-whisper} shows the duty cycle in the upper 
plot and the reliability in the lower plot.
We find that \name reduces the duty cycle by a factor of two for all updates $u$.
Considering only the energy consumption during communication~(i.e., 
20~mA) and assuming a battery with a capacity of 2000~mAh, Crystal 
with \name increases the nodes' lifetime by a factor of 2.3 compared 
to ``standard'' Crystal (278~days vs. 119~days of lifetime) in this 
experiment\footnote{
	Crystal (u=0) at 0~dBm transmission power has an average duty cycle of 3.46\%.
	Accordingly the battery lasts for 120~days ($2000~\text{mAh}/(3.46\% \cdot 20~\text{mA} \cdot 24~\text{h})$).
	Crystal w/ \name (coll.) (u=0) at the same transmission power has a duty cycle of 1.18\% and thus, the battery lasts for 353~days ($2000~\text{mAh}/(1.18\% \cdot 20~\text{mA} \cdot 24~\text{h})$).
	As a result, in case of no data transmission (i.e., $u=0$) \name used in Crystal increases the network lifetime by a factor of 2.9 compared to ``standard'' Crystal.
	Considering $u=5$, i.e., among the 27 nodes in FlockLab, there are 5 data packets to transmit during an epoch, the duty cycle of Crystal (u=5) is 3.49\% and the duty cycle of Crystal w/ \name (coll.) (u=5) is 1.5\%.
	Using the computation described above, Crystal with \name has a 2.3 
	times higher network lifetime compared to Crystal without \name.
}.
The factor of how much \name increases the network lifetime depends on the size of the packets sent in the T-slot as well as the number of retransmissions.
A large packet requires a long duration of the T-slot compared to short packets.
The longer the duration of the T-slot, the higher is the gain in terms of energy-efficiency that \name is able to achieve.
%
\newline
However, one can also observe that Crystal with \name has a lower reliability compared to \enquote{standard} Crystal (e.g., 95.7\% vs. 98.4\% at -10~dBm with $u=5$).
Indeed, when the sink misses a \packlet in a \name slot, it keeps its radio turned off during the T-slot and thus, also misses the data packet. 
Thus, \name requires high reliability when used as wake-up service.
New sampling strategies or a higher value for $N_{tx}$, as shown in 
Section~\ref{sec:eval_microbenchmark_ntx}, can help to increase 
\name's~reliability.

\section{Related work}
\label{sec:related_work}

The overall architecture of \name builds on the concepts introduced by \glossy. 
The novel design elements that we introduce -- in particular packlets and direction-aware sampling -- make \name significantly more efficient than \glossy, especially for small payload sizes. 
\name's superior performance is obtained by completely eliminating gaps between consecutive, synchronous transmissions. 

Lim et al.~\cite{Lim2017+Competition} modify \glossy so that a packet is transmitted multiple times after a single reception. 
Consecutive packets are, however, not sent back-to-back as in \name 
but have gaps between them.
This is because they are transmitted as individual packets and, thus, the radio must still perform a turnaround even between consecutive transmissions. 
This increases the overall transmit time and strongly limits the use of sampling strategies as introduced in \name. 
Furthermore, due to the instability of the DCO, the alignment of concurrently transmitted packets decreases quickly with the number of packets. 
In \name, instead, the use of Flock~\cite{Brachmann2017+Flock} and the concept of packlets guarantee that transmissions are precisely aligned. 

Approaches that exploit scheduled \glossy floods to provide 
high-level protocols -- e.g., LWB~\cite{Ferrari2012+LWB}, 
Crystal~\cite{Istomin2016+Crystal,Istomin2018+Crystal2}, or 
LaneFlood~\cite{Brachmann2016+LaneFlood} -- could replace \glossy 
with \name to achieve a more efficient operation. 
Other, more complex protocols like Splash~\cite{Doddavenkatappa2013+Splash} or Pando~\cite{Du2015+Pando} could also benefit from integrating \name's design in their architecture.

Several authors proposed protocols to use some form of frequency diversity to make synchronous transmissions more robust to interference. 
These include full-fledged protocols like Splash~\cite{Doddavenkatappa2013+Splash} or Pando~\cite{Du2015+Pando} but also improved versions of \glossy like the one proposed by Sommer and Pignolet~\cite{Sommer2016+GlossyHopping}. 
While we have not yet implemented the use of multiple channels within \name, this is part of our future work. 


Other approaches related to \name are those that provide -- or can be used to implement -- a network-wide wake-up service.
Some basic techniques like Low-Power Listening~\cite{Polastre2004+BMAC} or Backcast~\cite{Dutta2008+Backcast} have been successfully used in Medium Access Control protocols to schedule nodes' rendezvous~\cite{Polastre2004+BMAC, Buettner2006+XMAC, Moss2008+BoX-MACs, Dutta2012+AMAC}.
They are, however, contention-based approaches and are inherently 
less performing -- both in terms of reliability and latency -- than 
approaches based on synchronous transmissions. 

Lastly, protocols that exploit wake-up radios -- like, e.g., Zippy~\cite{Sutton2015+Zippy}, ALBA-WUR~\cite{Spenza2015+ALBA-WUR}, or the approach proposed in~\cite{Liu2016+Network-Wide-Wakeup} -- are orthogonal to ours because they rely on specialized hardware to be available on network nodes.

\section{Conclusions}
\label{sec:conclusion}


This paper introduces \name, a novel primitive to provide quick and reliable network floods.
\name exploits synchronous transmissions as in \glossy but eliminates any gap between consecutive transmissions of the packet to flood. 
This allows \name to halve the radio-on time of the nodes with respect to \glossy while maintaining a comparable or even higher reliability -- as demonstrated through our experiments on the FlockLab testbed. 
\name can be used as a stand-alone primitive to disseminate small values or be integrated in place of \glossy in protocols like, e.g., Crystal~\cite{Istomin2016+Crystal}. 
The source code of \name is publicly available at \small{\url{http://github.com/martinabr/whisper}}.

\section*{Acknowledgements}
\label{sec:ack}


The authors would like to thank Timofei Istomin for his technical support with the implementation of Crystal and the anonymous reviewers for the precious feedback. 
	This work was supported by the ERCIM Alain Bensoussan postdoc fellowship program, the Swedish Research Council VR through the ChaosNet and AgreeOnIT projects, and the Swedish Foundation for Strategic Research SSF through the LoWi project and the Smart Implicit Interaction project (RIT15-0046). The work described in this paper was performed while the first author was a PhD student at the Faculty of Computer Science of TU Dresden.

	
	\balance
	\bibliographystyle{ACM-Reference-Format}
	\bibliography{bib/sigproc}  


\begin{thebibliography}{00}


\ifx \showCODEN    \undefined \def \showCODEN     #1{\unskip}     \fi
\ifx \showDOI      \undefined \def \showDOI       #1{{\tt DOI:}\penalty0{#1}\ }
  \fi
\ifx \showISBNx    \undefined \def \showISBNx     #1{\unskip}     \fi
\ifx \showISBNxiii \undefined \def \showISBNxiii  #1{\unskip}     \fi
\ifx \showISSN     \undefined \def \showISSN      #1{\unskip}     \fi
\ifx \showLCCN     \undefined \def \showLCCN      #1{\unskip}     \fi
\ifx \shownote     \undefined \def \shownote      #1{#1}          \fi
\ifx \showarticletitle \undefined \def \showarticletitle #1{#1}   \fi
\ifx \showURL      \undefined \def \showURL       #1{#1}          \fi
\providecommand\bibfield[2]{#2}
\providecommand\bibinfo[2]{#2}
\providecommand\natexlab[1]{#1}

\bibitem[\protect\citeauthoryear{{Arago Systems}}{{Arago Systems}}{2011}]%
        {WiSMote}
\bibfield{author}{\bibinfo{person}{{Arago Systems}}.}
  \bibinfo{year}{2011}\natexlab{}.
\newblock \bibinfo{title}{WiSMote datasheet}.
\newblock   (\bibinfo{year}{2011}).
\newblock


\bibitem[\protect\citeauthoryear{{Atmel}}{{Atmel}}{2009}]%
        {AT86RF231}
\bibfield{author}{\bibinfo{person}{{Atmel}}.} \bibinfo{year}{2009}\natexlab{}.
\newblock \bibinfo{title}{AT86RF231 datasheet}.
\newblock   (\bibinfo{year}{2009}).
\newblock


\bibitem[\protect\citeauthoryear{{Atmel}}{{Atmel}}{2016}]%
        {AT86RF215}
\bibfield{author}{\bibinfo{person}{{Atmel}}.} \bibinfo{year}{2016}\natexlab{}.
\newblock \bibinfo{title}{AT86RF215 datasheet}.
\newblock   (\bibinfo{year}{2016}).
\newblock


\bibitem[\protect\citeauthoryear{Boano, Z{\'{u}}{\~{n}}iga, Brown, Roedig,
  Keppitiyagama, and R{\"{o}}mer}{Boano et~al\mbox{.}}{2014}]%
        {Boano2014+TempLab}
\bibfield{author}{\bibinfo{person}{Carlo~Alberto Boano}, \bibinfo{person}{Marco
  Z{\'{u}}{\~{n}}iga}, \bibinfo{person}{James Brown}, \bibinfo{person}{Utz
  Roedig}, \bibinfo{person}{Chamath Keppitiyagama}, {and} \bibinfo{person}{Kay
  R{\"{o}}mer}.} \bibinfo{year}{2014}\natexlab{}.
\newblock \showarticletitle{TempLab: A Testbed Infrastructure to Study the
  Impact of Temperature on Wireless Sensor Networks}. In
  \bibinfo{booktitle}{{\em Proceedings of the Conference on Information
  Processing in Sensor Networks (ACM/IEEE IPSN)}}. \bibinfo{pages}{95--106}.
\newblock


\bibitem[\protect\citeauthoryear{Brachmann, Landsiedel, and Santini}{Brachmann
  et~al\mbox{.}}{2016}]%
        {Brachmann2016+LaneFlood}
\bibfield{author}{\bibinfo{person}{Martina Brachmann}, \bibinfo{person}{Olaf
  Landsiedel}, {and} \bibinfo{person}{Silvia Santini}.}
  \bibinfo{year}{2016}\natexlab{}.
\newblock \showarticletitle{Concurrent Transmissions for Communication
  Protocols in the Internet of Things}. In \bibinfo{booktitle}{{\em Proceedings
  of the Conference on Local Computer Networks (IEEE LCN)}}.
  \bibinfo{pages}{406--414}.
\newblock


\bibitem[\protect\citeauthoryear{Brachmann, Landsiedel, and Santini}{Brachmann
  et~al\mbox{.}}{2017}]%
        {Brachmann2017+Flock}
\bibfield{author}{\bibinfo{person}{Martina Brachmann}, \bibinfo{person}{Olaf
  Landsiedel}, {and} \bibinfo{person}{Silvia Santini}.}
  \bibinfo{year}{2017}\natexlab{}.
\newblock \showarticletitle{Keep the Beat: On-The-Fly Clock Offset Compensation
  for Synchronous Transmissions in Low-Power Networks}. In
  \bibinfo{booktitle}{{\em Proceedings of the Conference on Local Computer
  Networks (IEEE LCN)}}. \bibinfo{pages}{303--311}.
\newblock


\bibitem[\protect\citeauthoryear{Buettner, Yee, Anderson, and Han}{Buettner
  et~al\mbox{.}}{2006}]%
        {Buettner2006+XMAC}
\bibfield{author}{\bibinfo{person}{Michael Buettner}, \bibinfo{person}{Gary~V.
  Yee}, \bibinfo{person}{Eric Anderson}, {and} \bibinfo{person}{Richard Han}.}
  \bibinfo{year}{2006}\natexlab{}.
\newblock \showarticletitle{X-MAC: A Short Preamble MAC Protocol for
  Duty-cycled Wireless Sensor Networks}. In \bibinfo{booktitle}{{\em
  Proceedings of the Conference on Embedded Networked Sensor Systems (ACM
  SenSys)}}. \bibinfo{pages}{307--320}.
\newblock


\bibitem[\protect\citeauthoryear{Doddavenkatappa, Chan, and
  Leong}{Doddavenkatappa et~al\mbox{.}}{2013}]%
        {Doddavenkatappa2013+Splash}
\bibfield{author}{\bibinfo{person}{Manjunath Doddavenkatappa},
  \bibinfo{person}{Mun~Choon Chan}, {and} \bibinfo{person}{Ben Leong}.}
  \bibinfo{year}{2013}\natexlab{}.
\newblock \showarticletitle{Splash: Fast Data Dissemination with Constructive
  Interference in Wireless Sensor Networks}. In \bibinfo{booktitle}{{\em
  Proceedings of the Symposium on Networked Systems Design \& Implementation
  (USENIX NSDI)}}. \bibinfo{pages}{269--282}.
\newblock


\bibitem[\protect\citeauthoryear{Du, Liando, Zhang, and Li}{Du
  et~al\mbox{.}}{2015}]%
        {Du2015+Pando}
\bibfield{author}{\bibinfo{person}{Wan Du}, \bibinfo{person}{Jansen~Christian
  Liando}, \bibinfo{person}{Huanle Zhang}, {and} \bibinfo{person}{Mo Li}.}
  \bibinfo{year}{2015}\natexlab{}.
\newblock \showarticletitle{When Pipelines Meet Fountain: Fast Data
  Dissemination in Wireless Sensor Networks}. In \bibinfo{booktitle}{{\em
  Proceedings of the Conference on Embedded Networked Sensor Systems (ACM
  SenSys)}}. \bibinfo{pages}{365--378}.
\newblock


\bibitem[\protect\citeauthoryear{Dutta, Dawson-Haggerty, Chen, Liang, and
  Terzis}{Dutta et~al\mbox{.}}{2012}]%
        {Dutta2012+AMAC}
\bibfield{author}{\bibinfo{person}{Prabal Dutta}, \bibinfo{person}{Stephen
  Dawson-Haggerty}, \bibinfo{person}{Yin Chen}, \bibinfo{person}{Chieh-Jan~Mike
  Liang}, {and} \bibinfo{person}{Andreas Terzis}.}
  \bibinfo{year}{2012}\natexlab{}.
\newblock \showarticletitle{A-MAC: A Versatile and Efficient Receiver-initiated
  Link Layer for Low-power Wireless}.
\newblock \bibinfo{journal}{{\em ACM Transactions on Sensor Networks\/}}
  \bibinfo{volume}{8} (\bibinfo{year}{2012}), \bibinfo{pages}{1--29}.
\newblock
Issue 4.


\bibitem[\protect\citeauthoryear{Dutta, Mus\u{a}loiu-E., Stoica, and
  Terzis}{Dutta et~al\mbox{.}}{2008}]%
        {Dutta2008+Backcast}
\bibfield{author}{\bibinfo{person}{Prabal Dutta}, \bibinfo{person}{R\u{a}zvan
  Mus\u{a}loiu-E.}, \bibinfo{person}{Ion Stoica}, {and}
  \bibinfo{person}{Andreas Terzis}.} \bibinfo{year}{2008}\natexlab{}.
\newblock \showarticletitle{Wireless ACK Collisions Not Considered Harmful}. In
  \bibinfo{booktitle}{{\em Proceedings of the Workshop on Hot Topics in
  Networks (ACM HotNets)}}. \bibinfo{pages}{1--6}.
\newblock


\bibitem[\protect\citeauthoryear{El-Hoiydi and Decotignie}{El-Hoiydi and
  Decotignie}{2004}]%
        {El-Hoiydi2004+WiseMac}
\bibfield{author}{\bibinfo{person}{Amre El-Hoiydi} {and}
  \bibinfo{person}{Jean~Dominique Decotignie}.}
  \bibinfo{year}{2004}\natexlab{}.
\newblock \showarticletitle{WiseMAC: An ultra low power MAC protocol for the
  downlink of infrastructure Wireless Sensor networks}. In
  \bibinfo{booktitle}{{\em Proceedings of the Symposium on Computers and
  Communications (IEEE ISCC)}}. \bibinfo{pages}{244--251}.
\newblock


\bibitem[\protect\citeauthoryear{Escobar, Moreno, Saez, Cabrera,
  Garcia-Jimenez, Cruz, Ruiz, Corona, Klaue, and Tati}{Escobar
  et~al\mbox{.}}{2018}]%
        {Escobar2018+Competition}
\bibfield{author}{\bibinfo{person}{Antonio Escobar}, \bibinfo{person}{Fernando
  Moreno}, \bibinfo{person}{Borja Saez}, \bibinfo{person}{Antonio~J. Cabrera},
  \bibinfo{person}{Javier Garcia-Jimenez}, \bibinfo{person}{Francisco~J. Cruz},
  \bibinfo{person}{Unai Ruiz}, \bibinfo{person}{Angel Corona},
  \bibinfo{person}{Jirka Klaue}, {and} \bibinfo{person}{Divya Tati}.}
  \bibinfo{year}{2018}\natexlab{}.
\newblock \showarticletitle{Competition: BigBangBus}. In
  \bibinfo{booktitle}{{\em Proceedings of the European Conference on Wireless
  Sensor Networks (EWSN)}}. \bibinfo{pages}{213--214}.
\newblock


\bibitem[\protect\citeauthoryear{Ferrari, Zimmerling, Mottola, and
  Thiele}{Ferrari et~al\mbox{.}}{2012}]%
        {Ferrari2012+LWB}
\bibfield{author}{\bibinfo{person}{Federico Ferrari}, \bibinfo{person}{Marco
  Zimmerling}, \bibinfo{person}{Luca Mottola}, {and} \bibinfo{person}{Lothar
  Thiele}.} \bibinfo{year}{2012}\natexlab{}.
\newblock \showarticletitle{Low-power wireless bus}. In
  \bibinfo{booktitle}{{\em Proceedings of the Conference on Embedded Networked
  Sensor Systems (ACM SenSys)}}. \bibinfo{pages}{1--14}.
\newblock


\bibitem[\protect\citeauthoryear{Ferrari, Zimmerling, Thiele, and
  Saukh}{Ferrari et~al\mbox{.}}{2011}]%
        {Ferrari2011+Glossy}
\bibfield{author}{\bibinfo{person}{Federico Ferrari}, \bibinfo{person}{Marco
  Zimmerling}, \bibinfo{person}{Lothar Thiele}, {and} \bibinfo{person}{Olga
  Saukh}.} \bibinfo{year}{2011}\natexlab{}.
\newblock \showarticletitle{Efficient network flooding and time synchronization
  with Glossy}. In \bibinfo{booktitle}{{\em Proceedings of the Conference on
  Information Processing in Sensor Networks (ACM/IEEE IPSN)}}.
  \bibinfo{pages}{73--84}.
\newblock


\bibitem[\protect\citeauthoryear{{IEEE Computer Society}}{{IEEE Computer
  Society}}{2016}]%
        {IEEE802.15.4}
\bibfield{author}{\bibinfo{person}{{IEEE Computer Society}}.}
  \bibinfo{year}{2016}\natexlab{}.
\newblock \bibinfo{title}{IEEE Standard for Low-Rate Wireless Networks}.
\newblock   (\bibinfo{year}{2016}).
\newblock


\bibitem[\protect\citeauthoryear{Istomin, Murphy, Picco, and Raza}{Istomin
  et~al\mbox{.}}{2016}]%
        {Istomin2016+Crystal}
\bibfield{author}{\bibinfo{person}{Timofei Istomin}, \bibinfo{person}{Amy~L.
  Murphy}, \bibinfo{person}{Gian~Pietro Picco}, {and} \bibinfo{person}{Usman
  Raza}.} \bibinfo{year}{2016}\natexlab{}.
\newblock \showarticletitle{Data Prediction + Synchronous Transmissions =
  Ultra-low Power Wireless Sensor Networks}. In \bibinfo{booktitle}{{\em
  Proceedings of the Conference on Embedded Networked Sensor Systems (ACM
  SenSys)}}. \bibinfo{pages}{83--95}.
\newblock


\bibitem[\protect\citeauthoryear{Istomin, Trobinger, Murphy, and Picco}{Istomin
  et~al\mbox{.}}{2018}]%
        {Istomin2018+Crystal2}
\bibfield{author}{\bibinfo{person}{Timofei Istomin}, \bibinfo{person}{Matteo
  Trobinger}, \bibinfo{person}{Amy~L. Murphy}, {and}
  \bibinfo{person}{Gian~Pietro Picco}.} \bibinfo{year}{2018}\natexlab{}.
\newblock \showarticletitle{Interference-Resilient Ultra-Low Power Aperiodic
  Data Collection}. In \bibinfo{booktitle}{{\em Proceedings of the Conference
  on Information Processing in Sensor Networks (ACM/IEEE IPSN)}}.
  \bibinfo{pages}{84--95}.
\newblock


\bibitem[\protect\citeauthoryear{K{\"{o}}nig and Wattenhofer}{K{\"{o}}nig and
  Wattenhofer}{2016}]%
        {Konig2016+Maintaining-CI}
\bibfield{author}{\bibinfo{person}{Michael K{\"{o}}nig} {and}
  \bibinfo{person}{Roger Wattenhofer}.} \bibinfo{year}{2016}\natexlab{}.
\newblock \showarticletitle{Maintaining Constructive Interference Using
  Well-Synchronized Sensor Nodes}. In \bibinfo{booktitle}{{\em Proceedings of
  the Conference Distributed Computing in Sensor Systems (DCOSS)}}.
  \bibinfo{pages}{206--215}.
\newblock


\bibitem[\protect\citeauthoryear{Landsiedel, Ferrari, and
  Zimmerling}{Landsiedel et~al\mbox{.}}{2013}]%
        {Landsiedel2013+Chaos}
\bibfield{author}{\bibinfo{person}{Olaf Landsiedel}, \bibinfo{person}{Federico
  Ferrari}, {and} \bibinfo{person}{Marco Zimmerling}.}
  \bibinfo{year}{2013}\natexlab{}.
\newblock \showarticletitle{Chaos: Versatile and Efficient All-to-All Data
  Sharing and In-Network Processing at Scale}. In \bibinfo{booktitle}{{\em
  Proceedings of the Conference on Embedded Networked Sensor Systems (ACM
  SenSys)}}. \bibinfo{pages}{1--14}.
\newblock


\bibitem[\protect\citeauthoryear{Liang, Priyantha, Liu, and Terzis}{Liang
  et~al\mbox{.}}{2010}]%
        {Liang2010+Multi-Headers}
\bibfield{author}{\bibinfo{person}{Chieh-Jan~Mike Liang},
  \bibinfo{person}{Nissanka~Bodhi Priyantha}, \bibinfo{person}{Jie Liu}, {and}
  \bibinfo{person}{Andreas Terzis}.} \bibinfo{year}{2010}\natexlab{}.
\newblock \showarticletitle{Surviving Wi-fi Interference in Low Power ZigBee
  Networks}. In \bibinfo{booktitle}{{\em Proceedings of the Conference on
  Embedded Networked Sensor Systems (ACM SenSys)}}. \bibinfo{pages}{309--322}.
\newblock


\bibitem[\protect\citeauthoryear{Lim, Ferrari, Zimmerling, Walser, Sommer, and
  Beutel}{Lim et~al\mbox{.}}{2013}]%
        {Lim2013+FlockLab}
\bibfield{author}{\bibinfo{person}{Roman Lim}, \bibinfo{person}{Federico
  Ferrari}, \bibinfo{person}{Marco Zimmerling}, \bibinfo{person}{Christoph
  Walser}, \bibinfo{person}{Philipp Sommer}, {and} \bibinfo{person}{Jan
  Beutel}.} \bibinfo{year}{2013}\natexlab{}.
\newblock \showarticletitle{FlockLab: A Testbed for Distributed, Synchronized
  Tracing and Profiling of Wireless Embedded Systems}. In
  \bibinfo{booktitle}{{\em Proceedings of the Conference on Information
  Processing in Sensor Networks (ACM/IEEE IPSN)}}. \bibinfo{pages}{153--166}.
\newblock


\bibitem[\protect\citeauthoryear{Lim, Forno, Sutton, and Thiele}{Lim
  et~al\mbox{.}}{2017}]%
        {Lim2017+Competition}
\bibfield{author}{\bibinfo{person}{Roman Lim}, \bibinfo{person}{Reto~Da Forno},
  \bibinfo{person}{Felix Sutton}, {and} \bibinfo{person}{Lothar Thiele}.}
  \bibinfo{year}{2017}\natexlab{}.
\newblock \showarticletitle{Competition: Robust Flooding using Back-to-Back
  Synchronous Transmissions with Channel-Hopping}. In \bibinfo{booktitle}{{\em
  Proceedings of the European Conference on Wireless Sensor Networks (EWSN)}}.
  \bibinfo{pages}{270--271}.
\newblock


\bibitem[\protect\citeauthoryear{Liu, Cao, Tang, and Wen}{Liu
  et~al\mbox{.}}{2016}]%
        {Liu2016+Network-Wide-Wakeup}
\bibfield{author}{\bibinfo{person}{Xuefeng Liu}, \bibinfo{person}{Jiannong
  Cao}, \bibinfo{person}{Shaojie Tang}, {and} \bibinfo{person}{Jiaqi Wen}.}
  \bibinfo{year}{2016}\natexlab{}.
\newblock \showarticletitle{Enabling Reliable and Network-Wide Wakeup in
  Wireless Sensor Networks}.
\newblock \bibinfo{journal}{{\em IEEE Transactions on Wireless
  Communications\/}}  \bibinfo{volume}{15} (\bibinfo{year}{2016}),
  \bibinfo{pages}{2262--2275}.
\newblock
Issue 3.


\bibitem[\protect\citeauthoryear{Ma, Zhang, Tang, Li, He, Zhang, Wei, and
  Theel}{Ma et~al\mbox{.}}{2018}]%
        {Ma2018+Competition}
\bibfield{author}{\bibinfo{person}{Xiaoyuan Ma}, \bibinfo{person}{Peilin
  Zhang}, \bibinfo{person}{Weisheng Tang}, \bibinfo{person}{Xin Li},
  \bibinfo{person}{Wangji He}, \bibinfo{person}{Fuping Zhang},
  \bibinfo{person}{Jianming Wei}, {and} \bibinfo{person}{Oliver Theel}.}
  \bibinfo{year}{2018}\natexlab{}.
\newblock \showarticletitle{Competition: Using Enhanced OF{\(\partial\)}COIN to
  Monitor Multiple Concurrent Events under Adverse Conditions}. In
  \bibinfo{booktitle}{{\em Proceedings of the European Conference on Wireless
  Sensor Networks (EWSN)}}. \bibinfo{pages}{211--212}.
\newblock


\bibitem[\protect\citeauthoryear{Mohammad, Doddavenkatappa, and Chan}{Mohammad
  et~al\mbox{.}}{2017}]%
        {Mohammad2017+Syncast}
\bibfield{author}{\bibinfo{person}{Mobashir Mohammad},
  \bibinfo{person}{Manjunath Doddavenkatappa}, {and} \bibinfo{person}{Mun~Choon
  Chan}.} \bibinfo{year}{2017}\natexlab{}.
\newblock \showarticletitle{Improving Performance of Synchronous
  Transmission-Based Protocols Using Capture Effect over Multichannels}.
\newblock \bibinfo{journal}{{\em ACM Transactions on Sensor Networks\/}}
  \bibinfo{volume}{13} (\bibinfo{year}{2017}), \bibinfo{pages}{1--26}.
\newblock
Issue 2.


\bibitem[\protect\citeauthoryear{Moss and Levis}{Moss and Levis}{2008}]%
        {Moss2008+BoX-MACs}
\bibfield{author}{\bibinfo{person}{David Moss} {and} \bibinfo{person}{Philip
  Levis}.} \bibinfo{year}{2008}\natexlab{}.
\newblock \bibinfo{booktitle}{{\em BoX-MACs: Exploiting Physical and Link Layer
  Boundariesin Low-Power Networking}}.
\newblock {T}echnical {R}eport. \bibinfo{pages}{1--12} pages.
\newblock


\bibitem[\protect\citeauthoryear{Nahas and Landsiedel}{Nahas and
  Landsiedel}{2018}]%
        {Nahas2018+Competition}
\bibfield{author}{\bibinfo{person}{Beshr~Al Nahas} {and} \bibinfo{person}{Olaf
  Landsiedel}.} \bibinfo{year}{2018}\natexlab{}.
\newblock \showarticletitle{Competition: Aggressive Synchronous Transmissions
  with In-network Processing for Dependable All-to-All Communication}. In
  \bibinfo{booktitle}{{\em Proceedings of the European Conference on Wireless
  Sensor Networks (EWSN)}}. \bibinfo{pages}{209--210}.
\newblock


\bibitem[\protect\citeauthoryear{{Olimex Ltd.}}{{Olimex Ltd.}}{2013}]%
        {Olimex2013+MSP430CCRF}
\bibfield{author}{\bibinfo{person}{{Olimex Ltd.}}}
  \bibinfo{year}{2013}\natexlab{}.
\newblock \bibinfo{title}{MSP430-CCRF development board}.
\newblock   (\bibinfo{year}{2013}).
\newblock
\newblock
\shownote{User's manual.}


\bibitem[\protect\citeauthoryear{Polastre, Hill, and Culler}{Polastre
  et~al\mbox{.}}{2004}]%
        {Polastre2004+BMAC}
\bibfield{author}{\bibinfo{person}{Joseph Polastre}, \bibinfo{person}{Jason
  Hill}, {and} \bibinfo{person}{David Culler}.}
  \bibinfo{year}{2004}\natexlab{}.
\newblock \showarticletitle{Versatile Low Power Media Access for Wireless
  Sensor Networks}. In \bibinfo{booktitle}{{\em Proceedings of the Conference
  on Embedded Networked Sensor Systems (ACM SenSys)}}.
  \bibinfo{pages}{95--107}.
\newblock


\bibitem[\protect\citeauthoryear{Saha and Chan}{Saha and Chan}{2017}]%
        {Saha2017+Many-to-one}
\bibfield{author}{\bibinfo{person}{Sudipta Saha} {and}
  \bibinfo{person}{Mun~Choon Chan}.} \bibinfo{year}{2017}\natexlab{}.
\newblock \showarticletitle{Design and Application of a Many-to-One
  Communication Protocol}. In \bibinfo{booktitle}{{\em Proceedings of the
  Conference on Computer Communications (IEEE INFOCOM)}}.
  \bibinfo{pages}{1--9}.
\newblock


\bibitem[\protect\citeauthoryear{{Semtech}}{{Semtech}}{2015}]%
        {SX1211}
\bibfield{author}{\bibinfo{person}{{Semtech}}.}
  \bibinfo{year}{2015}\natexlab{}.
\newblock \bibinfo{title}{SX1211 datasheet}.
\newblock   (\bibinfo{year}{2015}).
\newblock


\bibitem[\protect\citeauthoryear{Sommer and Pignolet}{Sommer and
  Pignolet}{2016}]%
        {Sommer2016+GlossyHopping}
\bibfield{author}{\bibinfo{person}{Philipp Sommer} {and}
  \bibinfo{person}{Yvonne-Anne Pignolet}.} \bibinfo{year}{2016}\natexlab{}.
\newblock \showarticletitle{Competition: Dependable Network Flooding Using
  Glossy with Channel-Hopping}. In \bibinfo{booktitle}{{\em Proceedings of the
  European Conference on Wireless Sensor Networks (EWSN)}}.
  \bibinfo{pages}{303--303}.
\newblock


\bibitem[\protect\citeauthoryear{Spenza, Magno, Basagni, Benini, Paoli, and
  Petrioli}{Spenza et~al\mbox{.}}{2015}]%
        {Spenza2015+ALBA-WUR}
\bibfield{author}{\bibinfo{person}{Dora Spenza}, \bibinfo{person}{Michele
  Magno}, \bibinfo{person}{Stefano Basagni}, \bibinfo{person}{Luca Benini},
  \bibinfo{person}{Mario Paoli}, {and} \bibinfo{person}{Chiara Petrioli}.}
  \bibinfo{year}{2015}\natexlab{}.
\newblock \showarticletitle{Beyond Duty Cycling: Wake-up Radio with Selective
  Awakenings for Long-lived Wireless Sensing Systems}. In
  \bibinfo{booktitle}{{\em Proceedings of the Conference on Computer
  Communications (IEEE INFOCOM)}}. \bibinfo{pages}{522--530}.
\newblock


\bibitem[\protect\citeauthoryear{Sutton, Buchli, Beutel, and Thiele}{Sutton
  et~al\mbox{.}}{2015}]%
        {Sutton2015+Zippy}
\bibfield{author}{\bibinfo{person}{Felix Sutton}, \bibinfo{person}{Bernhard
  Buchli}, \bibinfo{person}{Jan Beutel}, {and} \bibinfo{person}{Lothar
  Thiele}.} \bibinfo{year}{2015}\natexlab{}.
\newblock \showarticletitle{Zippy: On-Demand Network Flooding}. In
  \bibinfo{booktitle}{{\em Proceedings of the Conference on Embedded Networked
  Sensor Systems (ACM SenSys)}}. \bibinfo{pages}{45--58}.
\newblock


\bibitem[\protect\citeauthoryear{{Texas Instruments}}{{Texas
  Instruments}}{2007}]%
        {CC25202007}
\bibfield{author}{\bibinfo{person}{{Texas Instruments}}.}
  \bibinfo{year}{2007}\natexlab{}.
\newblock \bibinfo{title}{CC2520 datasheet}.
\newblock   (\bibinfo{year}{2007}).
\newblock


\bibitem[\protect\citeauthoryear{{Texas Instruments}}{{Texas
  Instruments}}{2011}]%
        {msp430f1611}
\bibfield{author}{\bibinfo{person}{{Texas Instruments}}.}
  \bibinfo{year}{2011}\natexlab{}.
\newblock \bibinfo{title}{MSP430F1611 datasheet}.
\newblock   (\bibinfo{year}{2011}).
\newblock


\bibitem[\protect\citeauthoryear{{Texas Instruments}}{{Texas
  Instruments}}{2013a}]%
        {CC11012013}
\bibfield{author}{\bibinfo{person}{{Texas Instruments}}.}
  \bibinfo{year}{2013}\natexlab{a}.
\newblock \bibinfo{title}{CC1101 datasheet}.
\newblock   (\bibinfo{year}{2013}).
\newblock


\bibitem[\protect\citeauthoryear{{Texas Instruments}}{{Texas
  Instruments}}{2013b}]%
        {CC24202013}
\bibfield{author}{\bibinfo{person}{{Texas Instruments}}.}
  \bibinfo{year}{2013}\natexlab{b}.
\newblock \bibinfo{title}{CC2420 datasheet}.
\newblock \url{http://www.ti.com/product/CC2420/}.   (\bibinfo{year}{2013}).
\newblock


\bibitem[\protect\citeauthoryear{Trobinger, Istomin, Murphy, and
  Picco}{Trobinger et~al\mbox{.}}{2018}]%
        {Trobinger2018+Competition}
\bibfield{author}{\bibinfo{person}{Matteo Trobinger}, \bibinfo{person}{Timofei
  Istomin}, \bibinfo{person}{Amy~L. Murphy}, {and} \bibinfo{person}{Gian~Pietro
  Picco}.} \bibinfo{year}{2018}\natexlab{}.
\newblock \showarticletitle{Competition: CRYSTAL Clear: Making Interference
  Transparent}. In \bibinfo{booktitle}{{\em Proceedings of the European
  Conference on Wireless Sensor Networks (EWSN)}}. \bibinfo{pages}{217--218}.
\newblock


\bibitem[\protect\citeauthoryear{Yuan and Hollick}{Yuan and Hollick}{2013}]%
        {Yuan2013+Talk}
\bibfield{author}{\bibinfo{person}{Dingwen Yuan} {and}
  \bibinfo{person}{Matthias Hollick}.} \bibinfo{year}{2013}\natexlab{}.
\newblock \showarticletitle{Let's Talk Together: Understanding Concurrent
  Transmission in Wireless Sensor Networks}. In \bibinfo{booktitle}{{\em
  Proceedings of the Conference on Local Computer Networks (IEEE LCN)}}.
  \bibinfo{pages}{219--227}.
\newblock


\bibitem[\protect\citeauthoryear{Yuan, Riecker, and Hollick}{Yuan
  et~al\mbox{.}}{2014}]%
        {Yuan2014+Sparkle}
\bibfield{author}{\bibinfo{person}{Dingwen Yuan}, \bibinfo{person}{Michael
  Riecker}, {and} \bibinfo{person}{Matthias Hollick}.}
  \bibinfo{year}{2014}\natexlab{}.
\newblock \showarticletitle{Making 'Glossy' Networks Sparkle: Exploiting
  Concurrent Transmissions for Energy Efficient, Reliable, Ultra-Low Latency
  Communication in Wireless Control Networks}. In \bibinfo{booktitle}{{\em
  Proceedings of the European Conference on Wireless Sensor Networks (EWSN)}}.
  \bibinfo{pages}{133--149}.
\newblock


\end{thebibliography}
	
	\newpage
	\section*{Appendix A: Theoretical radio-on time for large payloads}

Table~\ref{tab:cmp_radio-on} shows the radio-on time for \glossy, \name, and \name~(lazy) for different payload sizes and network diameters. The cells highlighted in gray indicate when the radio-on time of Whisper (lazy) exceeds the one of Glossy. Thus, the non-highlighted cells in the columns for \name and \name~(lazy) show when they have achieved a lower radio-on time than \glossy.

\enlargethispage{10pt}

\small
\begin{table}[!htbp]
    \centering
    \caption{\emph{Theoretical radio-on time of \glossy, \name, and \name (lazy) for different payload sizes and hop counts.}
            \name has a lower radio-on time compared to \glossy.
            However, depending on the payload size and the number of hops, the radio-on time for \name (lazy) may exceed the one of \glossy (the gray marked cells).
            The values include a RX/TX turnaround time of 192~$\mu$s  for \glossy but neglect the software delay.
    }
    \addtolength{\tabcolsep}{-4pt}
    {\footnotesize \begin{tabular}{c *{6}{c} | *{6}{c} | *{6}{c}}
                \toprule
                \emph{Payload} & \multicolumn{6}{c}{\emph{\glossy}} & \multicolumn{6}{c}{\emph{Whisper}} & \multicolumn{6}{c}{\emph{\name (lazy)}} \\
                \emph{size} & \emph{Hop 1} & \emph{2} & \emph{3} & \emph{4} & \emph{5} & \emph{6} & \emph{Hop 1} & \emph{2} & \emph{3} & \emph{4} & \emph{5} & \emph{6} & \emph{Hop 1} & \emph{2} & \emph{3} & \emph{4} & \emph{5} & \emph{6} \\
                \emph{[byte]} & \emph{[ms]} & \emph{[ms]} & \emph{[ms]} & \emph{[ms]} & \emph{[ms]} & \emph{[ms]} & \emph{[ms]} & \emph{[ms]} & \emph{[ms]} & \emph{[ms]} & \emph{[ms]} & \emph{[ms]} & \emph{[ms]} & \emph{[ms]} & \emph{[ms]} & \emph{[ms]} & \emph{[ms]} & \emph{[ms]} \\
                \midrule
                1 & 2.30 & 2.72 & 3.14 & 3.55 & 3.97 & 4.38 & 1.12 & 1.34 & 1.34 & 1.34 & 1.34 & 1.34 & 1.12 & 1.57 & 2.02 & 2.46 & 2.91 & 3.36
                \\
                2 & 2.50 & 2.94 & 3.39 & 3.84 & 4.29 & 4.74 & 1.28 & 1.54 & 1.54 & 1.54 & 1.54 & 1.54 & 1.28 & 1.79 & 2.30 & 2.82 & 3.33 & 3.84
                \\
                3 & 2.69 & 3.17 & 3.65 & 4.13 & 4.61 & 5.09 & 1.44 & 1.73 & 1.73 & 1.73 & 1.73 & 1.73 & 1.44 & 2.02 & 2.59 & 3.17 & 3.74 & 4.32
                \\
                4 & 2.88 & 3.39 & 3.90 & 4.42 & 4.93 & 5.44 & 1.60 & 1.92 & 1.92 & 1.92 & 1.92 & 1.92 & 1.60 & 2.24 & 2.88 & 3.52 & 4.16 & 4.80
                \\
                5 & 3.07 & 3.62 & 4.16 & 4.70 & 5.25 & 5.79 & 1.76 & 2.11 & 2.11 & 2.11 & 2.11 & 2.11 & 1.76 & 2.46 & 3.17 & 3.87 & 4.58 & 5.28
                \\
                6 & 3.26 & 3.84 & 4.42 & 4.99 & 5.57 & 6.14 & 1.92 & 2.30 & 2.30 & 2.30 & 2.30 & 2.30 & 1.92 & 2.69 & 3.46 & 4.22 & 4.99 & 5.76
                \\
                7 & 3.46 & 4.06 & 4.67 & 5.28 & 5.89 & 6.50 & 2.08 & 2.50 & 2.50 & 2.50 & 2.50 & 2.50 & 2.08 & 2.91 & 3.74 & 4.58 & 5.41 & 6.24
                \\
                8 & 3.65 & 4.29 & 4.93 & 5.57 & 6.21 & 6.85 & 2.24 & 2.69 & 2.69 & 2.69 & 2.69 & 2.69 & 2.24 & 3.14 & 4.03 & 4.93 & 5.82 & 6.72
                \\
                9 & 3.84 & 4.51 & 5.18 & 5.86 & 6.53 & 7.20 & 2.40 & 2.88 & 2.88 & 2.88 & 2.88 & 2.88 & 2.40 & 3.36 & 4.32 & 5.28 & 6.24 & 7.20
                \\
                10 & 4.03 & 4.74 & 5.44 & 6.14 & 6.85 & 7.55 & 2.56 & 3.07 & 3.07 & 3.07 & 3.07 & 3.07 & 2.56 & 3.58 & 4.61 & 5.63 & 6.66 & \cellcolor{lightgray}7.68
                \\
                11 & 4.22 & 4.96 & 5.70 & 6.43 & 7.17 & 7.90 & 2.72 & 3.26 & 3.26 & 3.26 & 3.26 & 3.26 & 2.72 & 3.81 & 4.90 & 5.98 & 7.07 & \cellcolor{lightgray}8.16
                \\
                12 & 4.42 & 5.18 & 5.95 & 6.72 & 7.49 & 8.26 & 2.88 & 3.46 & 3.46 & 3.46 & 3.46 & 3.46 & 2.88 & 4.03 & 5.18 & 6.34 & 7.49 & \cellcolor{lightgray}8.64
                \\
                13 & 4.61 & 5.41 & 6.21 & 7.01 & 7.81 & 8.61 & 3.04 & 3.65 & 3.65 & 3.65 & 3.65 & 3.65 & 3.04 & 4.26 & 5.47 & 6.69 & \cellcolor{lightgray}7.90 & \cellcolor{lightgray}9.12
                \\
                14 & 4.80 & 5.63 & 6.46 & 7.30 & 8.13 & 8.96 & 3.20 & 3.84 & 3.84 & 3.84 & 3.84 & 3.84 & 3.20 & 4.48 & 5.76 & 7.04 & \cellcolor{lightgray}8.32 & \cellcolor{lightgray}9.60
                \\
                15 & 4.99 & 5.86 & 6.72 & 7.58 & 8.45 & 9.31 & 3.36 & 4.03 & 4.03 & 4.03 & 4.03 & 4.03 & 3.36 & 4.70 & 6.05 & 7.39 & \cellcolor{lightgray}8.74 & \cellcolor{lightgray}10.08
                \\
                16 & 5.18 & 6.08 & 6.98 & 7.87 & 8.77 & 9.66 & 3.52 & 4.22 & 4.22 & 4.22 & 4.22 & 4.22 & 3.52 & 4.93 & 6.34 & 7.74 & \cellcolor{lightgray}9.15 & \cellcolor{lightgray}10.56
                \\
                17 & 5.38 & 6.30 & 7.23 & 8.16 & 9.09 & 10.02 & 3.68 & 4.42 & 4.42 & 4.42 & 4.42 & 4.42 & 3.68 & 5.15 & 6.62 & 8.10 & \cellcolor{lightgray}9.57 & \cellcolor{lightgray}11.04
                \\
                18 & 5.57 & 6.53 & 7.49 & 8.45 & 9.41 & 10.37 & 3.84 & 4.61 & 4.61 & 4.61 & 4.61 & 4.61 & 3.84 & 5.38 & 6.91 & 8.45 & \cellcolor{lightgray}9.98 & \cellcolor{lightgray}11.52
                \\
                19 & 5.76 & 6.75 & 7.74 & 8.74 & 9.73 & 10.72 & 4.00 & 4.80 & 4.80 & 4.80 & 4.80 & 4.80 & 4.00 & 5.60 & 7.20 & \cellcolor{lightgray}8.80 & \cellcolor{lightgray}10.40 & \cellcolor{lightgray}12.00
                \\
                20 & 5.95 & 6.98 & 8.00 & 9.02 & 10.05 & 11.07 & 4.16 & 4.99 & 4.99 & 4.99 & 4.99 & 4.99 & 4.16 & 5.82 & 7.49 & \cellcolor{lightgray}9.15 & \cellcolor{lightgray}10.82 & \cellcolor{lightgray}12.48
                \\
                21 & 6.14 & 7.20 & 8.26 & 9.31 & 10.37 & 11.42 & 4.32 & 5.18 & 5.18 & 5.18 & 5.18 & 5.18 & 4.32 & 6.05 & 7.78 & \cellcolor{lightgray}9.50 & \cellcolor{lightgray}11.23 & \cellcolor{lightgray}12.96
                \\
                22 & 6.34 & 7.42 & 8.51 & 9.60 & 10.69 & 11.78 & 4.48 & 5.38 & 5.38 & 5.38 & 5.38 & 5.38 & 4.48 & 6.27 & 8.06 & \cellcolor{lightgray}9.86 & \cellcolor{lightgray}11.65 & \cellcolor{lightgray}13.44
                \\
                23 & 6.53 & 7.65 & 8.77 & 9.89 & 11.01 & 12.13 & 4.64 & 5.57 & 5.57 & 5.57 & 5.57 & 5.57 & 4.64 & 6.50 & 8.35 & \cellcolor{lightgray}10.21 & \cellcolor{lightgray}12.06 & \cellcolor{lightgray}13.92
                \\
                24 & 6.72 & 7.87 & 9.02 & 10.18 & 11.33 & 12.48 & 4.80 & 5.76 & 5.76 & 5.76 & 5.76 & 5.76 & 4.80 & 6.72 & 8.64 & \cellcolor{lightgray}10.56 & \cellcolor{lightgray}12.48 & \cellcolor{lightgray}14.40
                \\
                25 & 6.91 & 8.10 & 9.28 & 10.46 & 11.65 & 12.83 & 4.96 & 5.95 & 5.95 & 5.95 & 5.95 & 5.95 & 4.96 & 6.94 & 8.93 & \cellcolor{lightgray}10.91 & \cellcolor{lightgray}12.90 & \cellcolor{lightgray}14.88
                \\
                26 & 7.10 & 8.32 & 9.54 & 10.75 & 11.97 & 13.18 & 5.12 & 6.14 & 6.14 & 6.14 & 6.14 & 6.14 & 5.12 & 7.17 & 9.22 & \cellcolor{lightgray}11.26 & \cellcolor{lightgray}13.31 & \cellcolor{lightgray}15.36
                \\
                27 & 7.30 & 8.54 & 9.79 & 11.04 & 12.29 & 13.54 & 5.28 & 6.34 & 6.34 & 6.34 & 6.34 & 6.34 & 5.28 & 7.39 & 9.50 & \cellcolor{lightgray}11.62 & \cellcolor{lightgray}13.73 & \cellcolor{lightgray}15.84
                \\
                28 & 7.49 & 8.77 & 10.05 & 11.33 & 12.61 & 13.89 & 5.44 & 6.53 & 6.53 & 6.53 & 6.53 & 6.53 & 5.44 & 7.62 & 9.79 & \cellcolor{lightgray}11.97 & \cellcolor{lightgray}14.14 & \cellcolor{lightgray}16.32
                \\
                29 & 7.68 & 8.99 & 10.30 & 11.62 & 12.93 & 14.24 & 5.60 & 6.72 & 6.72 & 6.72 & 6.72 & 6.72 & 5.60 & 7.84 & 10.08 & \cellcolor{lightgray}12.32 & \cellcolor{lightgray}14.56 & \cellcolor{lightgray}16.80
                \\
                30 & 7.87 & 9.22 & 10.56 & 11.90 & 13.25 & 14.59 & 5.76 & 6.91 & 6.91 & 6.91 & 6.91 & 6.91 & 5.76 & 8.06 & 10.37 & \cellcolor{lightgray}12.67 & \cellcolor{lightgray}14.98 & \cellcolor{lightgray}17.28
                \\
                31 & 8.06 & 9.44 & 10.82 & 12.19 & 13.57 & 14.94 & 5.92 & 7.10 & 7.10 & 7.10 & 7.10 & 7.10 & 5.92 & 8.29 & 10.66 & \cellcolor{lightgray}13.02 & \cellcolor{lightgray}15.39 & \cellcolor{lightgray}17.76
                \\
                32 & 8.26 & 9.66 & 11.07 & 12.48 & 13.89 & 15.30 & 6.08 & 7.30 & 7.30 & 7.30 & 7.30 & 7.30 & 6.08 & 8.51 & 10.94 & \cellcolor{lightgray}13.38 & \cellcolor{lightgray}15.81 & \cellcolor{lightgray}18.24
                \\
                33 & 8.45 & 9.89 & 11.33 & 12.77 & 14.21 & 15.65 & 6.24 & 7.49 & 7.49 & 7.49 & 7.49 & 7.49 & 6.24 & 8.74 & 11.23 & \cellcolor{lightgray}13.73 & \cellcolor{lightgray}16.22 & \cellcolor{lightgray}18.72
                \\
                34 & 8.64 & 10.11 & 11.58 & 13.06 & 14.53 & 16.00 & 6.40 & 7.68 & 7.68 & 7.68 & 7.68 & 7.68 & 6.40 & 8.96 & 11.52 & \cellcolor{lightgray}14.08 & \cellcolor{lightgray}16.64 & \cellcolor{lightgray}19.20
                \\
                35 & 8.83 & 10.34 & 11.84 & 13.34 & 14.85 & 16.35 & 6.56 & 7.87 & 7.87 & 7.87 & 7.87 & 7.87 & 6.56 & 9.18 & 11.81 & \cellcolor{lightgray}14.43 & \cellcolor{lightgray}17.06 & \cellcolor{lightgray}19.68
                \\
                36 & 9.02 & 10.56 & 12.10 & 13.63 & 15.17 & 16.70 & 6.72 & 8.06 & 8.06 & 8.06 & 8.06 & 8.06 & 6.72 & 9.41 & 12.10 & \cellcolor{lightgray}14.78 & \cellcolor{lightgray}17.47 & \cellcolor{lightgray}20.16
                \\
                37 & 9.22 & 10.78 & 12.35 & 13.92 & 15.49 & 17.06 & 6.88 & 8.26 & 8.26 & 8.26 & 8.26 & 8.26 & 6.88 & 9.63 & \cellcolor{lightgray}12.38 & \cellcolor{lightgray}15.14 & \cellcolor{lightgray}17.89 & \cellcolor{lightgray}20.64
                \\
                38 & 9.41 & 11.01 & 12.61 & 14.21 & 15.81 & 17.41 & 7.04 & 8.45 & 8.45 & 8.45 & 8.45 & 8.45 & 7.04 & 9.86 & \cellcolor{lightgray}12.67 & \cellcolor{lightgray}15.49 & \cellcolor{lightgray}18.30 & \cellcolor{lightgray}21.12 \\[-1ex]
                \multicolumn{19}{c}{$\cdots$} \\[-1ex]
                124 & 25.92 & 30.27 & 34.62 & 38.98 & 43.33 & 47.68 & 20.8 & 24.96 & 24.96 & 24.96 & 24.96 & 24.96 & 20.8 & 29.12 & \cellcolor{lightgray}37.44 & \cellcolor{lightgray}45.76 & \cellcolor{lightgray}54.08 & \cellcolor{lightgray}62.4 
                \\
                125 & 26.11 & 30.5 & 34.88 & 39.26 & 43.65 & 48.03 & 20.96 & 25.15 & 25.15 & 25.15 & 25.15 & 25.15 & 20.96 & 29.34 & \cellcolor{lightgray}37.73 & \cellcolor{lightgray}46.11 & \cellcolor{lightgray}54.5 & \cellcolor{lightgray}62.88  \\
                \bottomrule
    \end{tabular}}
    \addtolength{\tabcolsep}{1pt}
    \label{tab:cmp_radio-on}
\end{table}

\end{document}